\documentclass[useAMS,usenatbib,a4paper,usegraphicx]{mn2e}
\usepackage{fixltx2e}       
\usepackage{aas_macros}     

\usepackage{txfonts}        

\usepackage{color}    
\usepackage{hyperref} 
\usepackage{ifthen}            
\usepackage[normalem]{ulem} 

\newcommand{\hrsim}{hMS}               
\newcommand{\lcdm}{$\Lambda$CDM}       

\newcommand{\vv}[1]{\bmath{#1}}        

\newcommand{\dotprod}{\mathbf{\cdot}}
\newcommand{\crossp}{\mathbf{\times}}

\newcommand{\DM}{\Delta\mu}                 

\renewcommand{\mp}{ m_\rmn{p} }
\newcommand{\Np}{N_\rmn{p}}
\newcommand{\Npin}{N_\rmn{p,inner}}

\newcommand{\jin}{j_\rmn{inner}}
\newcommand{\jsc}{\tilde{\jmath}}
\newcommand{\jscin}{\tilde{\jmath}_\rmn{inner}}

\newcommand{\Mh}{M_\rmn{h}}
\newcommand{\Rvir}{R_\rmn{vir}}

\newcommand{\Rhalfm}{R_{1/2}}
\newcommand{\tlb}{t_\rmn{lb}}  

\newcommand{\figw}{80 mm}     
\newcommand{\figwsml}{57 mm}  
\newcommand{\figwlrg}{150 mm} 

\newcommand{\Msol}{\rmn{M_{\sun}}}   
\newcommand{\munit}{ \, h^{-1} \Msol }  
\newcommand{\OmegaM}{\Omega_\rmn{M}}

\newcommand{\OmegaMzero}{\Omega_{\rmn{M}0}}
\newcommand{\OmegaLzero}{\Omega_{\Lambda0}}
\newcommand{\Omegabzero}{\Omega_{\rmn{b}0}}

\newcommand{\rhoc}{\rho_\rmn{c}}  
\newcommand{\rhoczero}{ \rho_{\rmn{c}0} }
\newcommand{\Deltac}{\Delta_\rmn{c}}  

\newcommand{\Mpc}{\rmn{Mpc}}            
\newcommand{\kpc}{\rmn{kpc}}            
\newcommand{\kms}{\rmn{km\,s^{-1}}}     
\newcommand{\Hunit}{ \, \kms \, \Mpc^{-1}} 
\newcommand{\lunit}{ \, h^{-1} \Mpc }      
\newcommand{\klunit}{\, h^{-1} \kpc }      
\newcommand{\Gyr}{\, \rmn{Gyr} }        
\newcommand{\Lbox}{ L_{\rmn{box}} }
\newcommand{\Npart}{N_{\rmn{part}} }

\newcommand{\fpath}{./} 

\defcitealias{paperII}{Paper II}

\newcommand{\annotate}{false}  
\newcommand{\dofigscripts}{false}
\newcommand{\drafting}{newver} 
\newcommand{\pbscrapsty}[1]{\textcolor{red}{\sout{#1}}}
\newcommand{\pbnewsty}[1]{\textbf{\textcolor{blue}{#1}}}

\ifthenelse{\equal{\annotate}{true}}{
  \newcommand{\pbtodo}[1]{ [\textbf{\textcolor{red}{#1}}] }
}{
  \newcommand{\pbtodo}[1]{}
}
\ifthenelse{\equal{\dofigscripts}{true}}{
  \newcommand{\pbscripts}[1]{ [\textbf{\texttt{\textcolor{magenta}{#1}}}] }
}{
  \newcommand{\pbscripts}[1]{}
}
\ifthenelse{\equal{\drafting}{oldver}}{      
  \newcommand{\pbscrap}[1]{#1}                  
  \newcommand{\pbnew}[1]{}                      
} {\ifthenelse{\equal{\drafting}{newver}}{   
    \newcommand{\pbscrap}[1]{}                  
    \newcommand{\pbnew}[1]{#1}                  
  } {
    \newcommand{\pbscrap}[1]{\pbscrapsty{#1}}   
    \newcommand{\pbnew}[1]{\pbnewsty{#1}}       
  }
}

\title[Spin flips I]{Spin flips I: Evolution of the angular momentum
  orientation of Milky Way-mass dark matter haloes}

\author[P. E. Bett \& C. S. Frenk]{Philip E. Bett$^{1}$\thanks{Email: p.e.bett@physics.org} and Carlos S. Frenk$^{2}$ \\
  $^{1}$Argelander-Institut f\"ur Astronomie, Universit\"at Bonn, 
  Auf dem H\"ugel 71, D-53121 Bonn, Germany\\
  $^{2}$Institute for Computational Cosmology, University of Durham,
  South Road, Durham, DH1 3LE, UK
}

\begin{document}
\date{\today}
\pagerange{\pageref{firstpage}--\pageref{lastpage}} \pubyear{2011}

\maketitle

\label{firstpage}

\begin{abstract}
  During the growth of a cold dark matter halo, the direction of its
  spin can undergo rapid changes. These could disrupt or even destroy
  a stellar disc forming in the halo, \pbnew{possibly resulting in the
    generation of a bulge or spheroid}. We investigate the frequency
  of significant changes in the orientation of the angular momentum
  vector of dark matter haloes (``spin flips''), and their degree of
  correlation with mergers. We focus on haloes of mass similar to that
  of the Milky Way (MW) halo at redshift $z=0$ ($\log_{10} M/\munit =
  12.0\rightarrow 12.5$) and consider flips in the spin of the whole
  halo or just its inner parts.  We find that a greater fraction of
  major mergers are associated with large spin flips than minor
  mergers. However, since major mergers are rare, the vast majority
  ($93\%$) of large whole-halo spin flips ($\theta \geq 45\degr$)
  coincide with \emph{small} mass changes, not major mergers.  The
  spin vector of the inner halo experiences much more frequent flips
  than the halo as a whole. Over their entire lifetime\pbnew{s}
  (i.e. after \pbnew{a} halo acquires half of its final
  mass), over $10\%$ of halos experience a flip of at least $45\degr$
  in the spin of the entire halo and nearly 60 percent experience a
  flip this large in the inner halo. These numbers are reduced to 9
  percent for the whole halo and 47 percent for the inner halo when we
  consider only haloes with no major mergers after formation.  Our
  analysis suggests that spin flips (whose effects are not currently
  included in galaxy formation models) could be an important factor in
  the morphological transformation of disc galaxies.
\end{abstract}


\section{Introduction}
The acquisition and evolution of angular momentum plays a central role
in the formation and evolution of cosmic structure.  Early work on the
acquisition of angular momentum by virialized matter clumps in a
cosmological context dates back to \cite{1951pca..conf..195H}, and the
development of the linear tidal torque theory
(\citealt{1969ApJ...155..393P, 1970Ap......6..320D,
  1970Afz.....6..581D, 1984ApJ...286...38W, 1996MNRAS.282..436C,
  1996MNRAS.282..455C}; see also \citealt{2002MNRAS.332..325P}, and
\citealt{2009IJMPD..18..173S}).  This approach starts to break down as
structure growth becomes non-linear \citep{1984ApJ...286...38W}, and
the subsequent evolution of dark matter halo angular momentum is
usually studied using $N$-body simulations.  This subject too has a
long and rich research history, with simulations improving in size and
resolution as the available computing power has increased
(e.g. \citealt{1971A&A....11..377P, 1979MNRAS.186..133E, defw85,
  1987ApJ...319..575B, 1988ApJ...327..507F, 1992ApJ...399..405W,
  1996MNRAS.281..716C}).  Recent simulations have established the
distribution of the angular momentum of dark matter haloes, and its
evolution, extremely accurately, from very large numbers of
well-resolved objects (e.g. \citealt{2001ApJ...555..240B,
  2005ApJ...634...51A, 2006ApJ...646..815S,
  2007MNRAS.381...41H,2007MNRAS.375..489H, bett07,bett10,
  Maccio2007,Maccio2008, 2008ApJ...678..621K, 2011MNRAS.411..584M}).

These studies have usually focused on the distribution and evolution
of the angular momentum \emph{magnitude}.  In contrast, the angular
momentum vector \emph{direction} is relatively less well studied, and
often only in terms of its orientation with respect to the halo shape
\citep[e.g.][]{1992ApJ...399..405W, 2005ApJ...627..647B,
  2006MNRAS.367.1781A, 2006ApJ...646..815S, 2007MNRAS.377...50H,
  bett07,bett10}, or with other structures on different scales, such
as galaxies \citep[e.g.][]{ 2002ApJ...576...21V, 2003MNRAS.346..177V,
  2003ApJ...597...35C, 2006PhRvD..74l3522G, 2009MNRAS.400...43C,
  2009ApJ...702.1250R, bett10, 2010ApJ...709.1321A,
  2010MNRAS.405..274H, 2011MNRAS.415.2607D}, or large-scale filaments
and voids \citep[e.g.][]{2005ApJ...627..647B, 2007MNRAS.381...41H,
  2007MNRAS.375..489H, 2010MNRAS.405..274H, 2007MNRAS.375..184B,
  2008MNRAS.389.1127P, 2008MNRAS.385..867C}.
\cite{2000MNRAS.311..762S} and \cite{2002MNRAS.332..325P} tracked the
\emph{Lagrangian} evolution of the mass in $z=0$ haloes, and showed
that their spin direction changes due to non-linear evolution, with
both the average deviation from the initial direction, and the scatter
in that angle, increasing with time.

The motivation for studying the angular momentum of haloes is the
influence it is believed to have on the formation and evolution of
galaxies. In today's cosmological paradigm, in which the dark matter
is a cold collisionless particle (CDM), haloes grow hierarchically,
through a series of mergers of ever larger objects.  In the basic
two-stage picture of galaxy formation
\citep{whiterees1978,1991ApJ...379...52W}, galaxies form and evolve
within these haloes, with the pattern of simple hierarchical growth
modified by the more complex physical processes available to the
baryons as they cycle between gas and stars.  It is usually assumed
that the gas and dark matter have the same initial distribution, and
thus the gas in a halo initially has the same angular momentum as the
halo itself.  The gas then collapses to form a (rotationally
supported) disc galaxy, conserving its angular momentum.  Thus, the
size of the galactic disc is directly related to the dark matter
halo's angular momentum \citep{1980MNRAS.193..189F,
  1998MNRAS.295..319M,Zavala_Frenk08}.  This basic picture is
frequently implemented in so-called `semi-analytic' models of galaxy
formation \citep{1991ApJ...379...52W}, in which the modelling of the
baryonic processes is grafted onto the merger histories of dark matter
haloes, derived either from an $N$-body simulation or constructed
analytically.  This approach has been recently reviewed by
\cite{2006RPPh...69.3101B} and \cite{2010PhR...495...33B}, and a
comparison of different models has been carried out by
\cite{2010MNRAS.406.1533D}.  It is important to emphasise that while
these models incorporate the size of the dark matter angular momentum
vector, they make no reference to its direction.

It has long been known that tidal forces can induce morphological
changes in galaxies \citep{1972ApJ...178..623T}. If the gravitational
potential varies significantly over a short timescale, a galactic disc
can be disrupted completely. Galaxy formation models thus assume that
a sufficiently big galaxy merger event will destroy a disc,
randomising the stellar orbits and forming a spheroid\footnote{Such a
  spheroid is often distinguished from a so-called pseudobulge
  \cite[e.g.][]{2004ARA&A..42..603K,2008IAUS..245....3F}.
  Pseudobulges are thought to form through secular evolution of the
  disc.}  \citep[e.g.][]{1977egsp.conf..401T, 1988ApJ...331..699B,
  1992ApJ...393..484B,1996ApJ...471..115B, 1992ApJ...400..460H,
  1993ApJ...409..548H}.  Indeed, this has been shown to occur in
merger simulations of individual objects
\citep[e.g.][]{2003ApJ...597..893N, 2005A&A...437...69B,
  2006ApJ...650..791C,2008MNRAS.384..386C}. The outcome of a merger
depends on the gas richness of the participants
\citep[e.g.][]{2008ApJ...683..597S, 2009ApJ...702..307S,
  2009ApJ...691.1168H, 2009MNRAS.397..802H, 2010ApJ...715..202H}, and
on the details of the star formation and feedback processes triggered
by the merger \citep[e.g.][]{2005MNRAS.363.1299O,
  Zavala_Frenk08,2009MNRAS.396..696S}.

In this paper, we consider the evolution of the {\it direction} of the
angular momentum vector (hereafter \emph{spin} direction, for brevity)
of dark matter haloes, a process that can affect the stability of a
disc within the halo. Sudden, large changes in the halo spin direction
are indicative of a significant disturbance to the halo. Such changes
would usually accompany a halo merger\footnote{In the case of a major
  merger, the resulting spin direction is correlated with the net
  orbital spin of the progenitors
  \citep{2005MNRAS.362.1099F}. Furthermore, there is a degree of
  correlation in the infall directions of satellite haloes
  (e.g. falling along filaments,
  \citealt{2004ApJ...603....7K,2005MNRAS.363..146L,
    2011MNRAS.413.3013L, 2011MNRAS.411.1525L}).}, which, in turn,
could result in a galaxy merger within the halo, and potentially the
destruction of an existing galactic disc (depending on the details of
the baryonic physics).  However, it is also possible for tidal forces
to disturb the halo \emph{without} there being an immediate merger --
for example, due to the flyby of a neighbouring halo. Recent work by
\cite{2011arXiv1103.1675S} has shown that halo flybys indeed occur
sufficiently frequently that they have a significant dynamical effect
on halo systems.  In such a situation, the spin direction could change
significantly, even if its magnitude does not. Such ``spin flips''
could have major consequences for the survival of a disc. The
disturbance in the internal structure of the halo itself could lead to
the distruction of the disc, as in a merger. However, it could also
torque the disc and change its spin direction without disrupting it
\citep{1989MNRAS.237..785O}, causing it instead to become misaligned
relative to the direction of new infalling material. In due course,
the accretion of misaligned material could lead to the disruption of
the disc and the formation of a spheroid. Such spin flips provide a
mechanism of spheroid formation that is not currently considered in
galaxy formation models.

$N$-body and hydrodynamical simulations have shown that large, rapid
changes to dark matter halo spin directions do indeed occur.
\cite{2005MNRAS.363.1299O} remarked that their simulated galaxy, which
had formed a small disc, flipped its orientation.  It then began to
accrete gas in a direction nearly perpendicular to the original disc
which was subsequently transformed into a bulge with a new disc later
forming\footnote{A preliminary analysis of this system, in the spirit
  of the present paper, can be seen in \cite{2010AIPC.1240..403B}.}.
\cite{2009ApJ...702.1250R} analysed haloes in simulations both with
and without baryons, at very high time resolution.  They found that,
although the spin magnitude changes by a factor of $\sim2$--$3$ after
a major merger, the orientation can change much more drastically, by
$\gtrsim 180\degr$. Furthermore, such large changes in angular
momentum orientation are not restricted to major mergers.  In the
simulations with baryons, the authors also found that the DM halo,
stellar disc, and gas component can often flip orientation with
respect to each other as the system evolves, even at late times where
there are few major mergers.  \cite{2009MNRAS.396..696S} found that
misalignment of a stellar disc with the accreting cold gas can
sometimes cause mass to transfer from the disc to a spheroidal
component, sometimes destroying the disc (and sometimes allowing a new
disc to be formed later).

In this \pbnew{and subsequent papers}, we use a dark
matter $N$-body simulation to assess the frequency of spin flip events
occurring in the lifetime of haloes. We do not model baryonic physics,
and instead concentrate on quantifying the amplitude and frequency of
spin flips. Our aim \pbnew{in this first paper} is to \pbnew{make an
  initial} assess\pbnew{ment of} the importance of spin flips as a
potential mechanism for the disruption of discs and the formation of
spheroids.  \pbnew{We} focus
\pbnew{here} on those haloes whose mass at $z=0$ is similar to that of
the Milky Way (MW)\pbnew{.}  \pbnew{We perform a
  more in-depth study on the distribution of spin flips in a
  subsequent paper (\citealt{paperII}, hereafter \citetalias{paperII}).}

The outline of this paper is as follows. In section \ref{s:MWanaly},
we describe the $N$-body simulation we use and our analysis procedure,
including details of halo identification, merger trees, halo
selection, and the definition of the quantities of interest here: the
fractional mass change, and the spin orientation change.  We present
our results in section \ref{s:MWresults}, describing both the joint
and cumulative distributions of spin flip and merger events, and
investigate the frequency of spin flips over \pbnew{the course of}
halo lifetimes.  We discuss our conclusions in section \ref{s:MWconc}.

\section{Simulation data and analysis}\label{s:MWanaly}
In this section we describe the $N$-body dark matter simulation we use
and the associated halo catalogues and merger trees constructed to
link each halo with its descendent in a subsequent output time. We
calculate various halo properties, and use the merger trees to
describe how these properties evolve over the lifetime of each halo.

\subsection{The hMS simulation, haloes and merger trees}
We use the \hrsim{} cosmological dark matter simulation. This was
carried out using the same \textsc{L-GADGET-2} code and \lcdm{}
cosmological parameters as the Millennium Simulation \citep{mill2005},
but with a smaller box size and higher resolution\footnote{The
  \hrsim{} simulation was previously used by \cite{neto07},
  \cite{2008MNRAS.387..536G}, \cite{2009MNRAS.398.1150B},
  \cite{2009MNRAS.399..550L}, and \cite{bett10}. }.  Relevant
simulation parameters are shown in Table~\ref{t:simparams}, while the
assumed cosmological parameters are shown in Table~\ref{t:cosparams}.
Throughout, we refer to cosmological density parameters $\Omega_i(z) =
\rho_i(z) / \rhoc(z)$, in terms of the mass density\footnote{The
  equivalent mass density of the cosmological constant, $\Lambda$, is
  $\rho_\Lambda = \Lambda c^2/ (8\pi G)$.} of component $i$ and the
critical density $\rhoc(z) = 3H^2(z) / (8\pi G)$, where $H(z)$ is the
Hubble parameter.  We use a subscript zero to denote parameters
evaluated at $z=0$, and parameterise the present day value of the
Hubble parameter as $H_0=100h \Hunit$.

\begin{table}
  \begin{center} 
    \begin{tabular}{ccr@{.}lr@{.}l}\hline
      $\Lbox$  & $\Npart$        & \multicolumn{2}{l}{$\mp$}     & \multicolumn{2}{l}{$\eta$}   \\
      $\lunit$ &                 & \multicolumn{2}{l}{$10^7\munit$} & \multicolumn{2}{l}{$\klunit$}\\
      $100$    & $729\times10^6$ & 9&518                            & 2&4                          \\
      \hline
    \end{tabular}
    \caption{Simulation parameters for the \hrsim{} simulation: box
      size, numbers and masses of particles, and gravitational
      softening $\eta$.}
    \label{t:simparams}
  \end{center}
\end{table}

\begin{table} 
  \begin{center}
    \begin{tabular}{c c c c c c}\hline
      $\OmegaLzero$ & $\OmegaMzero$ & $\Omegabzero$ & $h$    & $n$   &$\sigma_8$ \\
      $0.75$    & $0.25$    & $0.045$   & $0.73$ & $1.0$ & $0.9$     \\
      \hline
    \end{tabular}
    \caption{Cosmological parameters (at $z=0$) for the \hrsim{}
      simulation used in this paper: cosmological density parameters
      $\Omega_{i0}$, the Hubble parameter $h$, the spectral index $n$,
      and $\sigma_8$ is the linear theory mass variance in spheres of
      radius $8\lunit$ at $z=0$. As with the Millennium Simulation,
      these parameters were chosen to be good matches to the results
      of the 2dF galaxy redshift survey
      \citep{2001MNRAS.328.1039C,2002MNRAS.337.1068P} and the first
      year results of the WMAP microwave background satellite
      \citep{wmap1cos2003}.}
    \label{t:cosparams}
  \end{center}
\end{table}

As with the Millennium Simulation, at each simulation snapshot
particle groups were identified on-the-fly according to the
friends-of-friends algorithm (FoF), with a linking length parameter of
$b=0.2$ \citep{defw85}.  Subsequently, self-bound substructures within
these groups were found using the \textsc{Subfind} algorithm
\citep{subfind2001}.  Finally, the progenitors and descendents of each
particle group were found, creating a `merger tree' structure allowing
haloes to be tracked over time.  The merger tree algorithm (and
associated halo definition) used is that described in
\cite{harker2006}, which was originally designed for use with the
\textsc{Galform} semi-analytic galaxy formation model and the
Millennium Simulation\footnote{In particular, they correspond to the
  \texttt{DHalo} tables in the Millennium Simulation database
  \citep{milldb}} \citep{helly2003,bowergalform2006}.  Since the
construction of haloes and mergers trees from particle groups in
discrete snapshots is essential for the current work, we will now
describe this process in more detail.

The preliminary set of haloes consists, at each snapshot, of the FoF
particle groups, which in turn contain the self-bound substructures
identified by \textsc{Subfind} (these include the main body of the
halo, plus less massive subhaloes), as well as so-called ``fuzz''
particles not gravitationally bound to any structure in the halo.  It
is well known that the purely spatial nature of the FoF algorithm
allows multiple objects to be linked together spuriously, in the sense
that although they are close together they might not necessarily be
physically connected. A common example is that of two close objects
linked with a tenuous bridge of particles, which the FoF algorithms
identifies as a single ``halo'' (see \citealt{bett07} for a detailed
discussion and comparison of the effect of groupfinders on the
measured halo angular momentum and related properties).

Including additional physical information -- e.g. gravitational
binding from \textsc{Subfind}, and temporal evolution from merger
trees -- allows the operational definition of a halo to be refined, to
match better our physical intuition of what a halo is, and thus, when
used in conjunction with semi-analytic models of galaxy formation, to
allow better comparisons with both observations and hydrodynamic
simulations.  \pbnew{Following \cite{2002ApJ...568...52W}, a}
 ``splitting'' algorithm is applied to the basic FoF halo
catalogues, whereby spuriously linked subhaloes are split off from
their original FoF groups and identified as separate haloes in their
own right. A subhalo is split off from its original FoF parent if it
satisfies at least one of the following conditions: (1) The distance
between the subhalo centre and the parent centre is more than twice
the half-mass radius of the parent; or (2) the subhalo still has more
than $75\%$ of the mass it had when it was last identified as a
separate halo.  This yields halo catalogues containing more objects
than the corresponding FoF catalogues.  \cite{bett07} showed that
these `merger tree haloes' are a great improvement on both the simple
FoF groups and groups found from a simple spherical overdensity
criterion.

Merger trees for these haloes are constructed by tracking the
particles that constitute the subhaloes between each snapshot,
starting at early times and continuing to redshift $z=0$.  Each halo
or subhalo can have at most one descendent in a later snapshot.  The
most bound $10\%$ of a subhalo's mass (or $10$ most bound particles if
that is more massive) is located in the next snapshot.  Occasionally,
these particles might no longer reside in a subhalo in the next
snapshot: the subhalo might have temporarily dropped below
\textsc{Subfind}'s detection limit, or might be passing through a
high density region and be interpreted as unbound matter around that
density peak.  In practice therefore, the next five snapshots are
scanned to find the earliest time when these particles are again in
subhaloes. The descendent subhalo is then identified as the subhalo
containing the largest number of those most bound particles.  The
descendent of a halo is identified as the halo whose most massive
substructure (i.e. the main self-bound halo component) is the
descendent of its own most massive substructure.

It is possible that a subhalo's mass ends up distributed between
two (or more) subhaloes in a subsequent snapshot.  While one will be
identified as the descendent, the other will be left as a separate
``orphan'' object without a progenitor.  This situation is known as a
de-merger, and is a physical effect separate to the splitting of
groups described above -- here, sets of particles physically end up in
separate objects as the simulation evolves.

The end result of this process is a catalogue of haloes (groups of
self-bound substructures) identified at each snapshot, with at most
one descendent and one or more progenitors. Each halo identified at
$z=0$ is the root of its own tree, which branches into many progenitor
haloes at preceding output times.  In this paper, we study the
evolution of properties of individual haloes that at $z=0$ have a mass
corresponding roughly to that of the Milky Way.  After identifying an
appropriate halo at $z=0$, we track its evolution back by finding its
most massive progenitor at the preceding snapshot, then finding the
most massive of \emph{that} halo's progenitors, and so on.

It is important to note that, just like the halo definition and galaxy
formation model, the halo merger tree algorithm is not by any means
unique -- even within a given $N$-body simulation. The halo merger
trees used for the ``MPA'' semi-analytic models of the Millennium
Simulation (e.g. \citealt{mill2005} and \citealt{2007MNRAS.375....2D})
in fact track the binding-energy-weighted mass in the \textsc{Subfind}
subhaloes. Other methods that use splitting/stitching algorithms
similar to the one used here include those by
e.g. \cite{2008MNRAS.386..577F,2009MNRAS.394.1825F},
\cite{2009ApJ...701.2002G}; see also \cite{2006ApJ...647..763M}.
\cite{2009A&A...506..647T} provide a recent detailed study of halo
definition and merger tree algorithms.

\subsection{Halo property catalogues}
Various properties of the haloes are computed at each output time.
Properties are computed in the centre-of-momentum frame of each halo,
and in physical rather than comoving coordinates.  Each halo consists
of a set of $\Np$ particles, with each particle $i$ having mass $m_i =
\mp$, position $\vv{x}_i$ and velocity $\vv{v}_i$. The halo mass is
therefore $\Mh= \sum_{i=1}^{\Np}m_i = \Np\mp$.  The halo centre is
taken to be the location of the gravitational potential minimum of its
most massive substructure, as found by \textsc{Subfind}.  We define an
approximate ``virial'' radius, $\Rvir$, for the halo\footnote{Note
  that we do not use $\Rvir$ as a boundary for our halo. However, it
  provides a useful scale for the physical halo size needed in other
  properties.} by growing a sphere from the halo centre and computing
the density of the halo particles within. We locate $\Rvir$ at the
radius at which the density enclosed drops below a certain threshold
value computed at that snapshot, $\rho_\rmn{h} = \Deltac(z) \rhoc(z)$.
The threshold overdensity with respect to critical, $\Deltac(z)$, is
determined from the spherical collapse model \citep{ECF96}, using the
fitting formula of \cite{BN98}:
\begin{equation}
  \label{e:virdens}
  \Deltac(z) = 18\pi^2  + 82\left(\OmegaM(z)-1\right) - 39\left(\OmegaM(z)-1\right)^2
\end{equation}
In the case of the flat \lcdm{} universe used here, $\OmegaM(z) =
\OmegaMzero a^{-3} / \chi(z)$ and $\rhoc(z) = \rhoczero \chi(z)$,
where the expansion factor $a = (1+z)^{-1}$ and we define $\chi(z) =
\OmegaMzero a^{-3} + \OmegaLzero$ for convenience.

We compute halo energies as in \cite{bett07,bett10}. The kinetic
energy of a halo is given by $T = \frac{1}{2} \sum_{i=1}^{\Np} m_i
\vv{v}_i^2$, and the potential energy is computed as a double sum over
a random sample of $1000$ particles, using the same smoothing kernel
for gravitational softening as in the simulation itself.  We only use
the energies for some broad selection criteria (described below), and
random sampling provides a good approximation.

In this paper, we are mostly interested in the halo angular momentum
vector, $\vv{J} = \sum_{i=1}^{\Np}m_i \vv{x}_i \crossp \vv{v}_i$.  We
also define an inner halo angular momentum, $\vv{J}_\rmn{inner}$,
using the particles within $r_\rmn{inner} = 0.25\Rvir$.\footnote{This
  is similar to \cite{bett10}, although there we defined
  $r_\rmn{inner} = 10^{-0.6}\Rvir \approx 0.25\Rvir$}

\subsection{Halo selection}\label{s:hsel}
We need to select haloes at each snapshot from which reliable
measurements of angular momentum can be made.  The halo has to be well
defined, that is, both well resolved (consisting of a sufficiently
large number of particles), and reasonably relaxed (close to being
virialised).  Furthermore, the angular momentum magnitude cannot be
too small: if the angular momentum vectors of most particles are in
opposite directions and cancel, then the net direction will be
dominated by very few particles and will not be robust.  We follow the
approach of \cite{bett10}, defining a scaled angular
momentum\footnote{Note that $\jsc$ is identical to the alternative
  spin parameter $\lambda'$ introduced by \cite{2001ApJ...555..240B},
  modulo a factor of $\sqrt{2}$.} $\jsc$ as the ratio of the specific
angular momentum $j$ to that of a single particle in a Keplerian
orbit, $\jsc = j/\sqrt{G \Mh \Rvir}$, along with the analogous
quantity for the inner halo $\jscin = \jin/\sqrt{G M_\rmn{inner}
  0.25\Rvir}$.  As a basic way of assessing virialisation, we compute
$Q= 2T/U +1$, which should be around zero for a virialised halo.  We
use the same critical values for selection as \cite{bett10}; in
particular, haloes that pass the following three criteria are
retained:
\begin{eqnarray}
  \Np            & \geq & 1000 \\  
  \left|Q\right| & \leq & 0.5  \\  
  \log_{10} \jsc & \geq & -1.5      
\end{eqnarray}
(When considering changes to the inner halo, the criteria for $\Np$
and $\jsc$ are replaced by equivalent ones for $\Npin$ and $\jscin$.)
\pbnew{Note that these selection criteria are applied to haloes
  \emph{separately at each given snapshot}, rather than once for their
  whole lifetime.  Haloes can be excluded at one timestep (including
  at $z=0$), but still retained for study at subsequent or preceding
  timesteps.}

In addition to the particle number cut above, we restrict our analysis
to haloes that at $z=0$ have masses similar to that of the Milky Way
halo. That is, we retain only haloes whose final mass is $10\,506\mp
\leq M_0 < 33\,224\mp$, equivalent to the mass range $12.0 \leq
\log_{10} (M_0/\munit) < 12.5 $ (see Table~\ref{t:simparams}). 

\begin{figure}
  \centering\includegraphics[width=\figw]{\fpath 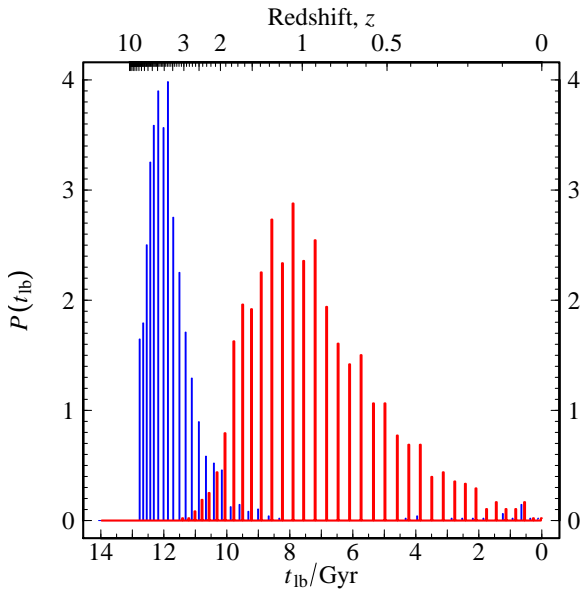}
  \caption{Histogram of formation times (red) and initial detection
    (`start') times (blue) for haloes that have MW masses and pass our
    \pbnew{standard} selection criteria at $z=0$, in
    terms of lookback time, $\tlb$, and redshift.  The formation and
    start times are computed using haloes at $z>0$ that pass just the
    $\Np$ selection criterion.  There is a histogram spike at each
    snapshot from $z<6.2$; the blue spikes are offset slightly to make
    them visible. \pbscripts{fig\_histo\_tform-tstart\_MWmass.R}}
   \label{f:MWformtimes}
\end{figure}

A visual inspection of the coevolution of different halo properties
for individual haloes suggested that a further two selection criteria
should also be applied.  Firstly, the early life of a halo is very
chaotic, with a high rate of mass accretion, mergers, and general
instability in halo properties.  So that our results are not dominated
by this early period, before the halo has properly formed (in some
sense), we restrict our analysis to the time period after the final
time when $M(z) < 0.5 M_0$ (where $M_0$ is the halo mass at $z=0$).
This corresponds to a commonly used simple definition of halo
``formation'' time \citep[e.g.][and references
therein]{1993MNRAS.262..627L, 2004MNRAS.349.1464S,
  2007MNRAS.381...41H, neto07, 2007MNRAS.376..977G,
  2008MNRAS.389.1419L}. \pbnew{Since stellar disks are unlikely to
  survive the early chaotic phase of halo formation \citep{Parry09},
  restricting attention to the period after the halo has formed is
  appropriate for investigating the frequency of spin flips that could
  alter the morphology of a disk galaxy. This condition must be borne
  in mind when interpreting the statistics that we present below.} We
plot the formation times of these haloes, according to this
definition, in Fig.~\ref{f:MWformtimes}.  The peak in the distribution
is around haloes forming at $z\approx 1$.

Secondly, we found that there are some occasions in which the halo
finder or merger tree algorithms make unphysical choices for which
subhaloes to incorporate into which haloes.  For example, if a
satellite halo was orbiting near the edge of a halo, then it might
``merge'' at one snapshot, then be identified as a separate object
again later, only to finally merge again afterwards.  During such
events, the angular momentum vector might appear to swing around
wildly, as a large mass at large radius would be added to, then
removed from, the total halo $\vv{J}$.  Since such changes are
\emph{not} due to a physical change in the halo angular momentum, but
instead are due to uncertainty in where to draw the halo boundary, we
should exclude such events.  It turns out that such events also cause
large changes in the halo kinetic energy $T$, as the bulk velocity of
the satellite halo will be incorporated into the main halo greatly
increasing its net velocity dispersion.  Thus, such events can be
identified by considering the arithmetic change in the virialisation
parameter $Q$; since the potential energy $U<0$ and does not change
much, \pbnew{an apparent sudden increase in $T$ makes} $Q$ 
appear to decrease suddenly.  By examining various cases, we chose to
exclude events\footnote{Since this effect is due to uncertainties in
  the halo boundary, we do not apply this exclusion criterion when
  considering the inner halo spin.} that have $\Delta Q \leq -0.3$.

Finally, we note that we analyse the halo population over the redshift
range $z<6.2$; in any case, the effects we describe will be most
visible at low redshift.  \pbnew{As shown in Fig.~\ref{f:MWformtimes},
  there are $46$ snapshots over this redshift range, with $23$ over
  the period $z<1$.}

\subsection{Evolution of halo properties}
Combining the merger tree data with the halo property catalogues at
each snapshot\pbnew{,} we can obtain the evolution of each halo property,
for each halo identified at $z=0$.  We are most interested in the
relationship between the \emph{change} in halo mass and the
\emph{change} in halo spin orientation, from snapshot to snapshot.
That is, we focus on two \emph{differential} halo properties, the
fractional mass change
\begin{equation}
  \label{e:DMfrac}
  \DM(t) := \frac{M(t) - M(t-\tau)}{M(t)},
\end{equation}
and the angular change in spin orientation
\begin{equation}
  \label{e:costheta}
  \cos\theta(t) := \frac{      \vv{J}(t)   \dotprod   \vv{J}(t-\tau)}
                        {\left|\vv{J}(t)\right| \left|\vv{J}(t-\tau)\right|},
\end{equation}
where $t$ is the time at which the quantity is measured, and $\tau$ is
the timescale over which we measure the halo property change; the
time $t-\tau$ precedes the time $t$.

In principle, we could simply look at the difference in halo
properties at adjacent snapshot times.  However, since the snapshots
in the \hrsim{} are not evenly spaced in time, this would not be a
fair way to analyse events in haloes at different times (the
intersnapshot time varies between $\sim 0.1$--$0.4\Gyr$ for $z\la
6$).  Instead, we choose a constant value for $\tau$, and simply
linearly interpolate the halo property in question between the values
at the snapshots before and after the time $t-\tau$.  The simulation
snapshots are in fact sufficiently closely spaced in time that this
interpolation is accurate.

We will refer to the property (or property change) of a given halo at
a given snapshot as an \emph{event}.  We shall use some fiducial
critical values to divide the distribution of events to aid
interpretation.  We shall consider a spin direction change of at least
$\theta_0=45\degr$ to be `large', and a fractional mass change of more
than $\DM_0=0.3$ to correspond to a major merger\footnote{$\DM$ is
only restricted to be $<1$; a value of $\DM=\frac{1}{3}$ means that
the mass has increased by $50\%$. If $\DM\ge \frac{1}{2}$, then the
halo has more than doubled in mass; we expect this to be rare, since
we are, by definition, comparing with the most massive progenitor.
Negative values of $\DM$ are possible, corresponding to mass
\emph{loss} between snapshots; $\DM=-1$ means that the halo has lost
$50\%$ of its previous mass.}.  For the sake of brevity, we shall
refer to events with $\DM \leq 0.3$ as minor mergers, even though they
could be smooth accretion (i.e. not the merging with a satellite
halo), or even mass loss.

The choice of event timescale is non\pbnew{-}trivial, since any
characteristic halo dynamic timescale is likely to depend on halo mass
and size (and therefore also on cosmology and time) -- but we wish to
use a single timescale for all haloes at all times, so that we can
compare events at different times in different haloes
fairly. \pbnew{The timescale we choose will therefore be only an
  approximation for the actual timescale of any particular halo.}

We consider the orbital timescale for a particle in a Keplerian orbit
at the half-mass radius \pbnew{$\Rhalfm$} of a \pbnew{model} halo
bounded by the radius enclosing the density $\Deltac(z)\rhoc(z)$.  For
a halo of a given mass, a concentration can be found by assuming an
NFW density profile \citep{nfw96,nfw97} and using the redshift
dependent mass-concentration relation of
\cite{2011MNRAS.411..584M}\pbnew{. The}  half-mass
radius for a halo of a given concentration can then be found using the
fitting formula of \cite{2001MNRAS.321..155L}, allowing a timescale to
be computed as
\begin{equation}
\tau_{1/2} = \sqrt{\frac{2\Rhalfm^3}{G M}}.
\end{equation}
For haloes in the mass range we consider in this paper, $\tau_{1/2}$
varies from about $0.37\Gyr$ at $z=1$ to about $0.63\Gyr$ at $z=0$.
(The values computed analytically using the fitting formulae outlined
above agree with those measured directly from haloes in the
simulation.)  We therefore take a fixed value for the event timescale
of $\tau=0.5\Gyr$, although we note that we do not expect our results
to depend qualitatively on the \pbnew{\emph{exact}} value
used\pbnew{; for some key results we show their
  dependence on $\tau$.}  We will consider the choice of $\tau$ in
more detail in \pbnew{\citetalias{paperII}.} 
It is important to note that $\tau$ is the timescale for our
measurements of halo changes, and the physical timescale of flips or
mergers can be much shorter.

Using the whole-halo selection criteria described in
section~\ref{s:hsel} gives us a population of $35\,279$ events.
 \pbnew{When we select instead} for the
inner halo spin, we have $29\,889$ events.


\section{Results}\label{s:MWresults}

\subsection{The distribution of flips and mergers}
We start by examining how changes in the spin orientation of the halo
correlate with changes in the halo mass that occur at the same time.

\subsubsection{Distribution of whole halo flip and merger
events}\label{s:MWflipdistro} 
\begin{figure} 
  \centering\includegraphics[width=\figw]{\fpath 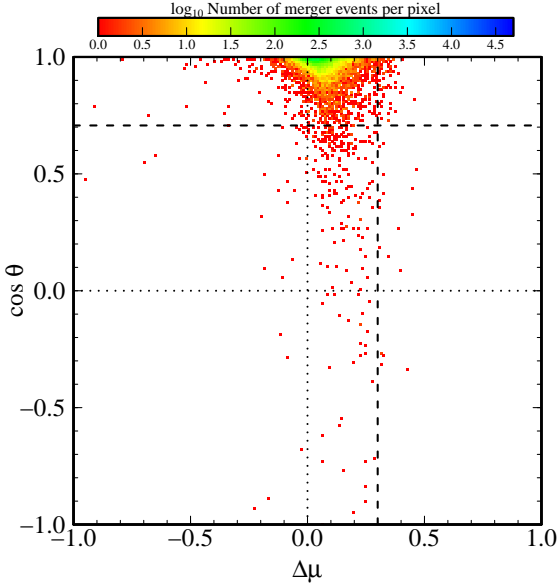}
  \caption{Distribution of events as a function of fractional mass
    change, $\DM$, and spin orientation change, $\cos\theta$.  Dotted
    lines mark the origin and dashed lines indicate our fiducial
    critical values for major mergers ($\DM\geq 0.3$) and large flips
    ($\theta>45\degr$). \pbscripts{fig\_DM-cos\_MWmass\_distroimg.R} }
  \label{f:MWflipdistro}
\end{figure}

The distribution of events for MW-final-mass haloes is shown in
Fig. \ref{f:MWflipdistro}.  There are very few major mergers or large
flips, with most events located around ``no change''
($\cos\theta\approx 1$, $\DM \approx 0$).  Most of the spread in
$\cos\theta$ is located between no mass change and our fiducial
threshold for major mergers ($\DM=0.3$)

\begin{figure} 
  \centering\includegraphics[width=\figw]{\fpath 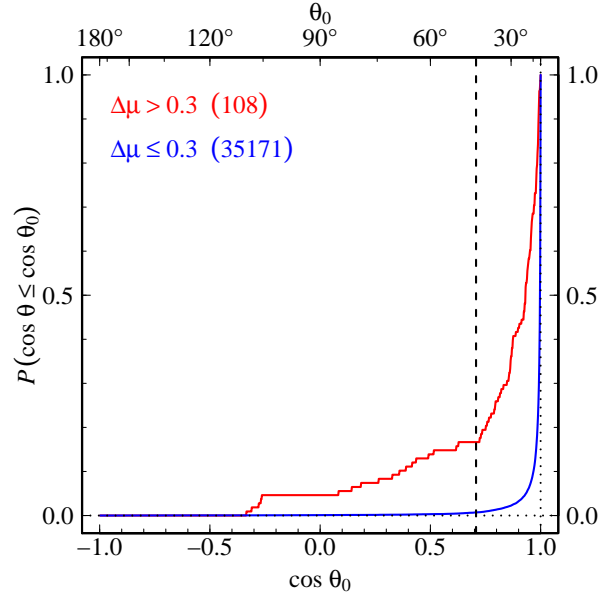}
  \caption{Cumulative distribution of events with spin misalignments
    of at least $\theta_0$ degrees.  The dashed line shows our
    fiducial value of $\theta_0=45\degr$.  We show results for a
    limiting merger fraction of $\DM_0=0.3$ (red: major mergers; blue:
    minor mergers). The number of events in each case is written in
    the legend.  \pbscripts{fig\_cdf\_cos\_MWmass.R} }
  \label{f:coscdf}
\end{figure}

Since we are interested in mergers and flips above and below some
critical value, rather than \emph{at} some value, it is useful to
examine the cumulative distribution functions (CDFs) of the data. The
CDF of $\cos\theta$ is shown in Fig. \ref{f:coscdf}.  We can see that,
if we consider just events without major mergers ($\DM\leq 0.3$), then
only a very small fraction have large flips: about $0.7\%$ have flips
of $45\degr$ or more.  If we select only major mergers, then since we
have now excluded the main peak of the distribution we find a much
higher proportion of events with large flips: about $17\%$ have flips
of at least $45\degr$ (although the major mergers themselves represent
only $0.3\%$ of the total event distribution).


\begin{figure} 
  \centering\includegraphics[width=\figw]{\fpath 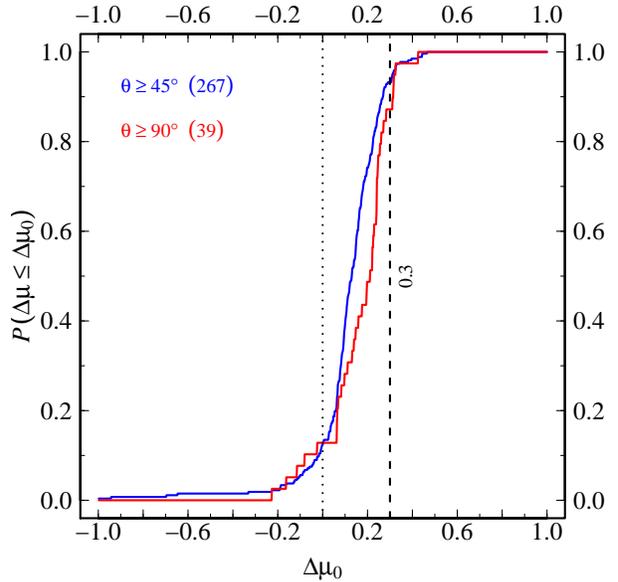}
  \caption{Cumulative distribution of events with fractional mass change
    of $\DM_0$ or less, for events with large spin flips (red: at
    least $45\degr$; blue: at least $90\degr$). The dashed line shows
    our fiducial value of $\DM_0=0.3$, and the dotted line marks no
    mass change.  \pbscripts{fig\_cdf\_dmunaligned\_MWmass.R}}
  \label{f:dmfraccdf}
\end{figure}

We can also consider the CDF of $\DM$ (Fig. \ref{f:dmfraccdf}).  In
this case, if we select just large flips, we find that the vast
majority ($93\%$ of those with $\theta \geq 45\degr$) coincide with
minor mergers ($\DM \leq 0.3$).  


\subsubsection{Distribution of flips of the inner spin}\label{s:MWjinner}
\begin{figure*}
  \centering
  \includegraphics[width=\figwsml]{\fpath 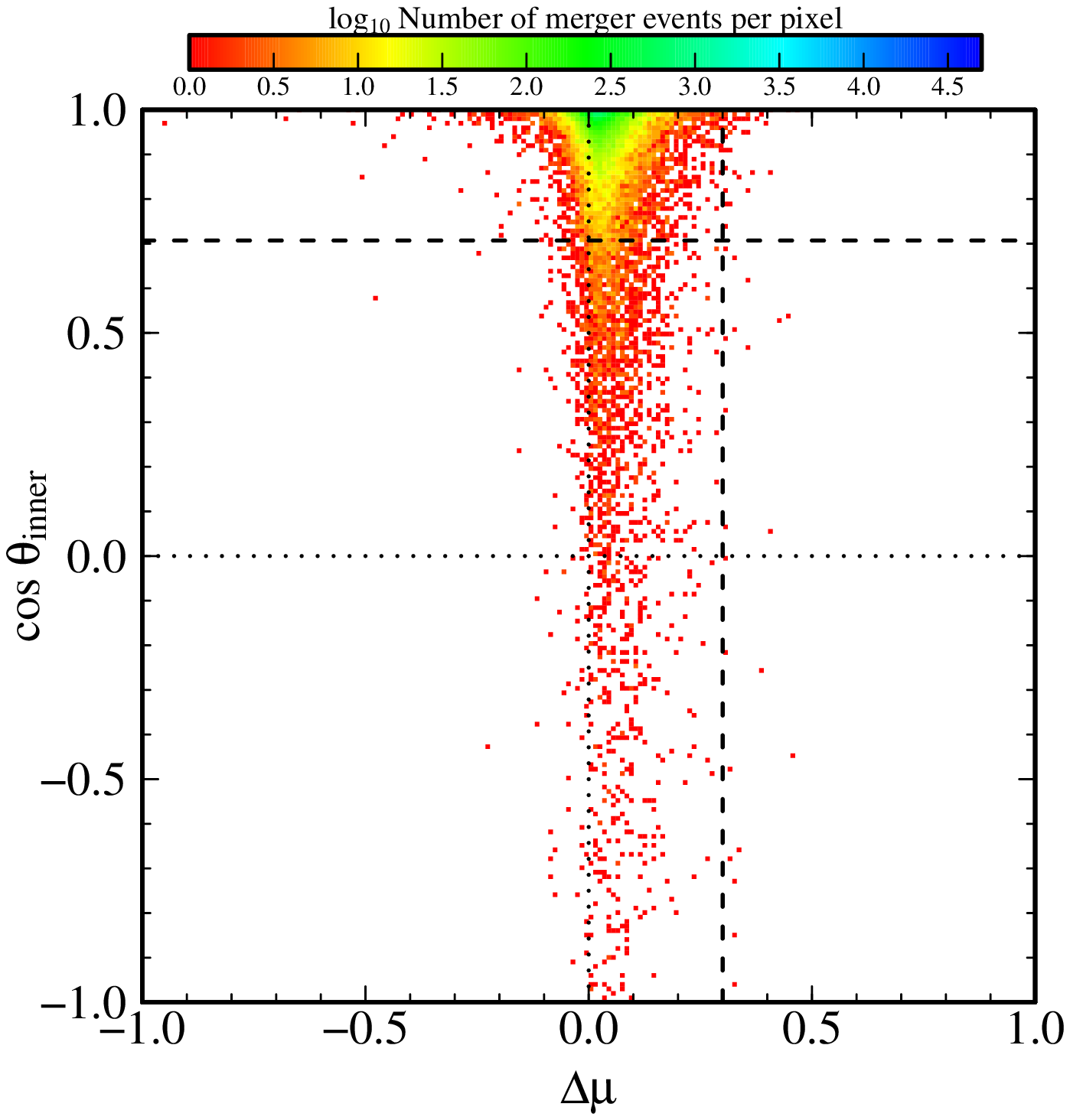}
  \includegraphics[width=\figwsml]{\fpath 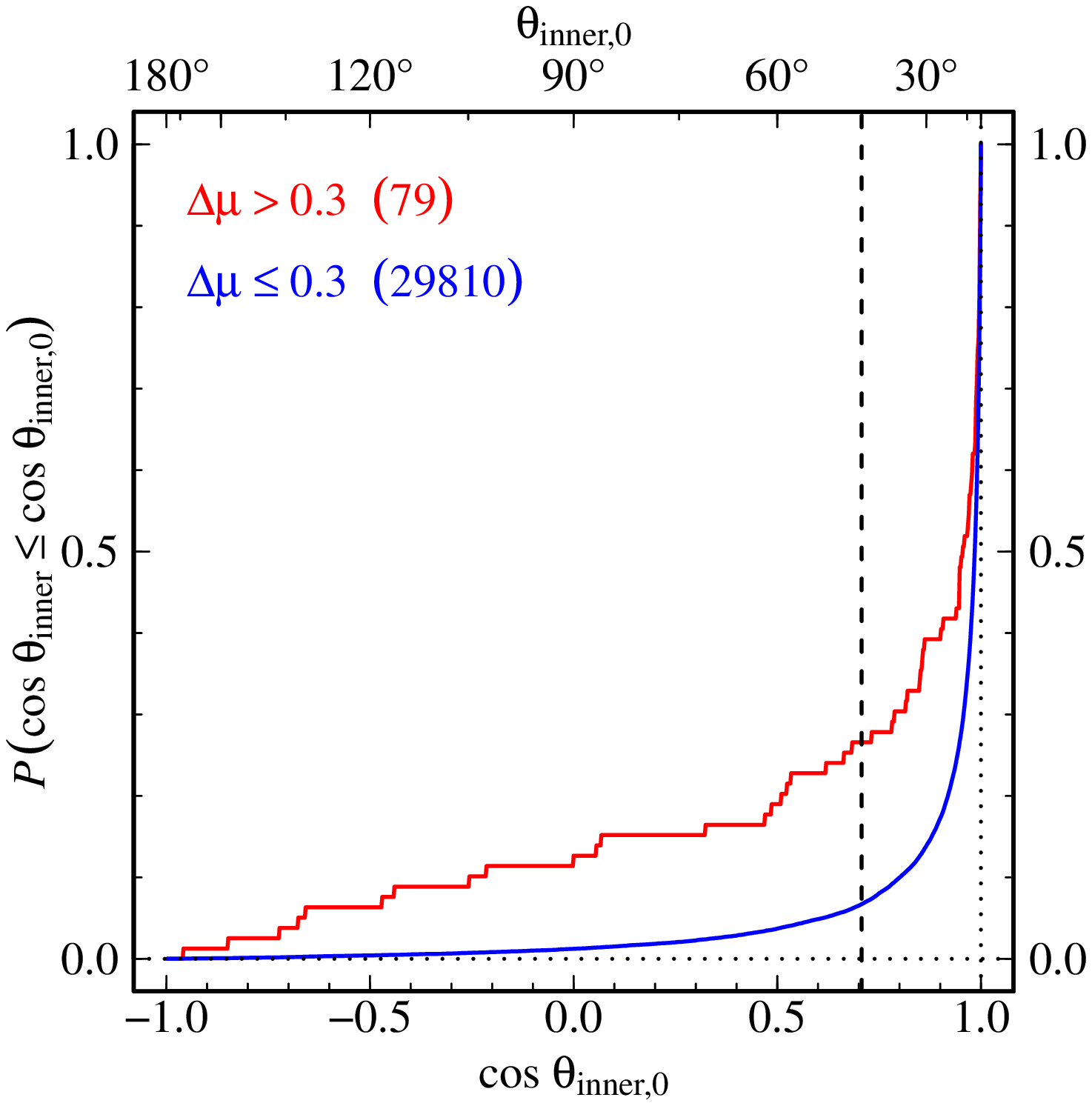}
  \includegraphics[width=\figwsml]{\fpath 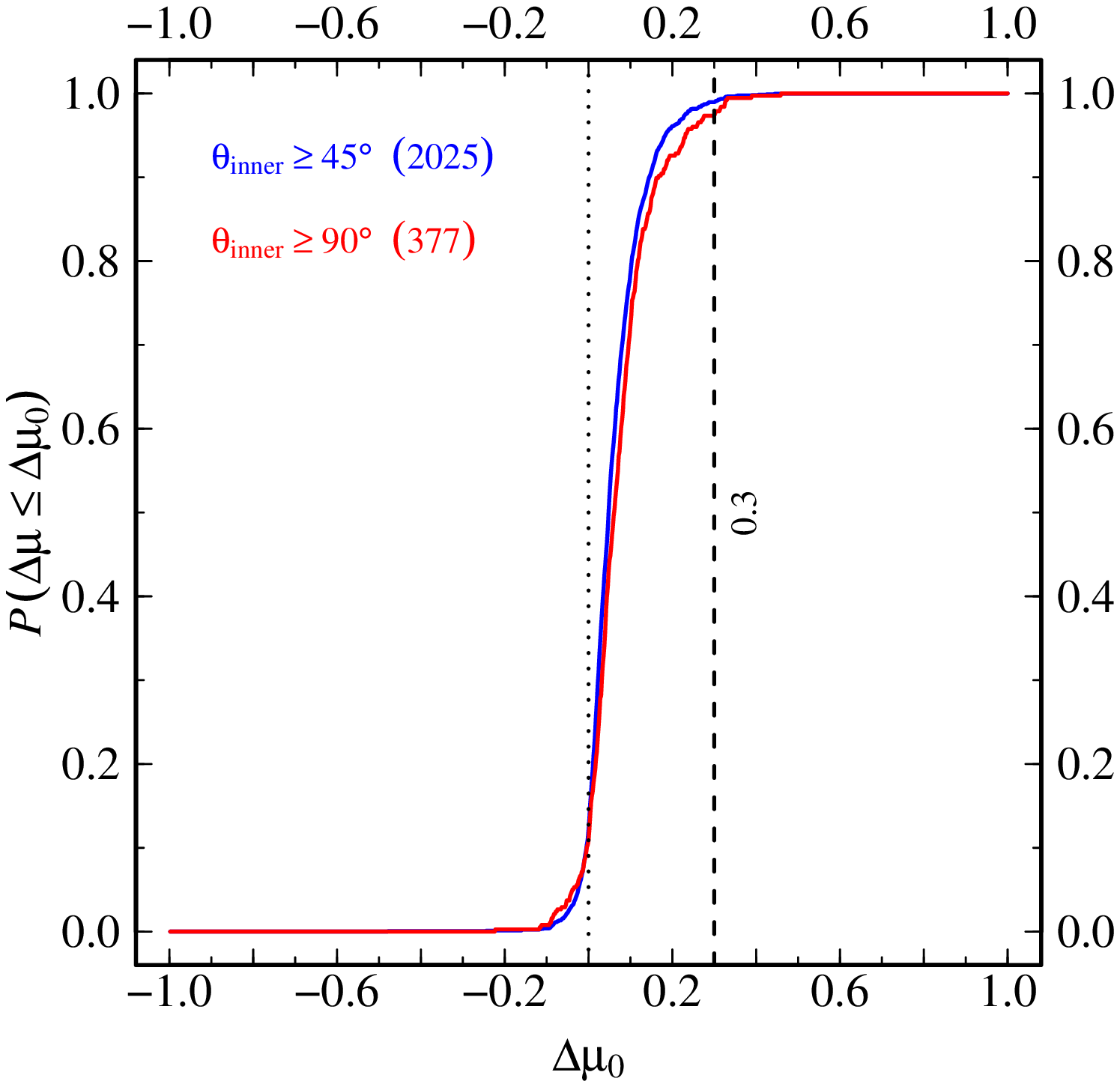}

  \caption{Left: event distribution as a function of the
  inner halo spin flip versus the total halo fractional mass change.
  Middle: Cumulative distribution function for merger events with
  inner spin misalignments of at least $\theta_{\rmn{inner,}0}$
  degrees, for major merger events (red) and minor mergers (blue).
  Right: Cumulative distribution of events with mass changes $\leq
  \DM_0$, for spin flips of at least $45\degr$ (blue) and at least
  $90\degr$ (red).  \pbscripts{fig\_DM-cosinner\_MWmass\_distroimg.R}
  \pbscripts{fig\_cdf\_cosinner\_MWmass.R}
  \pbscripts{fig\_cdf\_dmunaligned\_cutoncosinner\_MWmass.R}}
  \label{f:MWinnerplots}
\end{figure*}

The distribution of flips of the inner halo angular momentum is more
directly relevant when considering the stability of galaxies that
might form within.  The \pbnew{joint} distribution of events as a
function of the inner halo spin direction change 
\pbnew{and} the mass change of the whole halo is shown in
Fig.~\ref{f:MWinnerplots}.  In comparison to the distribution for
total halo spin flips, the inner halo exhibits a far greater spread to
low-$\cos\theta$.

Cumulative distributions are shown in the middle and right panels of
Fig.~\ref{f:MWinnerplots}.  We find that the frequency of minor merger
events (the blue line in the middle panel) that have a large inner
spin flip is about $6.7\%$, which is a significant increase on that
for total halo flips shown in Fig.~\ref{f:coscdf} ($0.7\%$).  The
fraction of major merger events that also have significant inner flips
is slightly increased, to $26.6\%$.  Selecting just large flips (right
panel), the frequencies are similarly increased compared to the total
halo flip distribution: $98.9\%$ of flips of at least $45\degr$
coincide with minor mergers, dropping slightly to $97.3\%$ for flips
of at least $90\degr$.



\subsection{Spin flips over halo lifetimes}
While it is important to understand the overall frequency of flip
events, and their tendency to correlate with mergers, we are also
concerned with the frequency of spin orientation changes over the
course of halo lifetimes.

\subsubsection{Flips and coincident mergers as a function of flip
  duration}\label{s:MWhflipfracDt}
An important question to answer is what is the likelihood of a halo
exhibiting a spin flip (of a given magnitude $\theta_0$ and measured
over a timescale $\tau$) at some point during its lifetime?  This can
be further specified by restricting attention to spin flips that do
(or do not) coincide with a major merger.

\begin{figure*} 
  \centering\includegraphics[width=\figwlrg]{\fpath 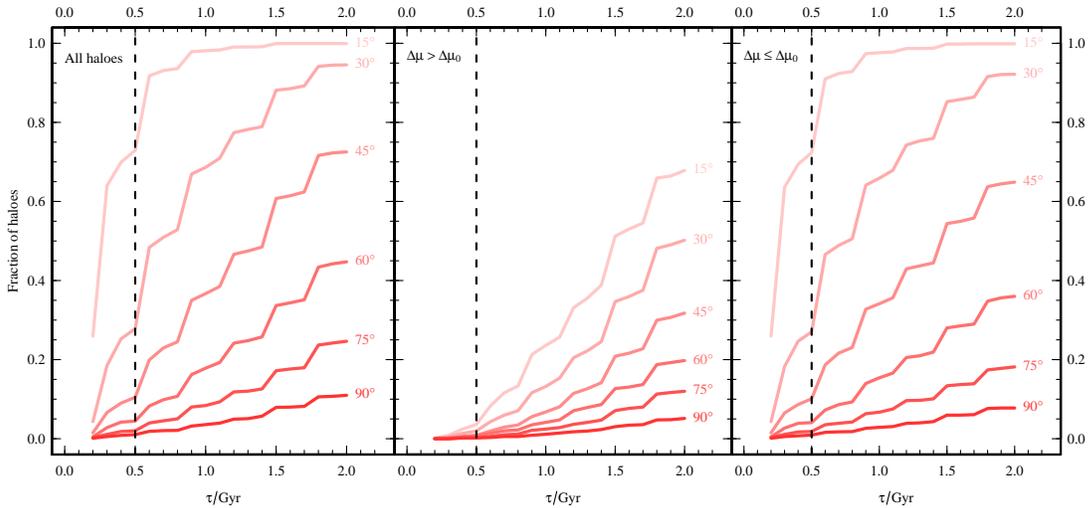}
  \caption{The fraction of haloes with Milky-Way masses at $z=0$ that
    have at least one flip of at least $\theta_0$ and duration $\tau$
    (left).  Middle: same, but adding the constraint that the flip
    must coincide with a major merger $\DM>\DM_0$, with $\DM_0=0.3$.
    Right: same, but the flip must \emph{not} coincide with a major
    merger. \pbscripts{fig\_halototflipprobsDt\_MWmass.R}}
  \label{f:MWhtotflipfracDt}
\end{figure*}

We answer these questions in Fig.~\ref{f:MWhtotflipfracDt}, for a
range of values of $\theta_0$ and $\tau$, given our fiducial major
merger threshold of $\DM_0=0.3$.  As one would expect, the likelihood
of getting a spin flip increases as one considers longer timescales or
flips of smaller magnitudes. (The steps are an artefact of the
discrete and irregular snapshot times, coupled with our interpolation
scheme and relatively small halo population.  Increasing $\tau$ causes
jumps as the snapshots used for interpolation change, and this occurs
at different values of $\tau$ for haloes in different snapshots. With
\pbnew{a larger halo sample,}, the lines become
smooth curves, which we demonstrate in \pbnew{\citetalias{paperII}.})

Quantitatively, considering flips using our fiducial values of
$\theta_0=45\degr$ and $\tau=0.5\Gyr$, we find that $10.5\%$ of
Milky-Way final mass haloes ($172$ haloes) experience such a flip at
some point in their lives.  If we consider just those for which such
flips coincide with major mergers, this drops to just $0.9\%$ ($14$
haloes).  We find that $10.1\%$ of haloes ($166$) experience a large
flip without a major merger.


\begin{figure*} 
  \centering\includegraphics[width=\figwlrg]{\fpath 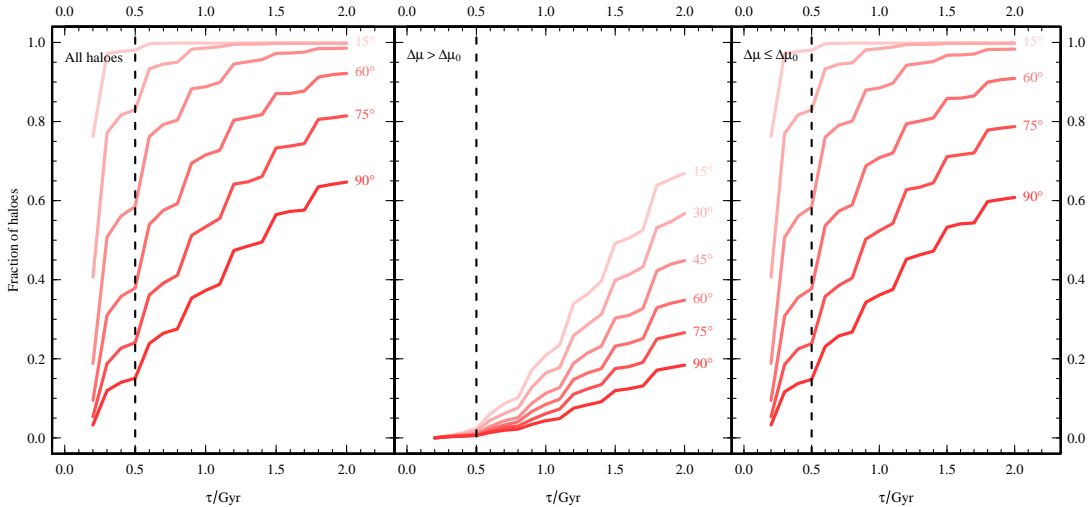}
  \caption{As Fig.~\ref{f:MWhtotflipfracDt}, but for flips in the
    inner halo angular momentum. }
  \label{f:MWhinnflipfracDt}
\end{figure*}

We can construct a similar plot for changes to the inner halo angular
momentum direction, which we show in Fig.~\ref{f:MWhinnflipfracDt}.
There is an increased tendency for haloes to have flips that do not
coincide with major mergers, for all flip angles $\theta_0$ and
timescales $\tau$.  For example, for our fiducial choice of $\tau$ and
$\theta_0$, we find that $58.5\%$ ($783$ haloes) have large flips at
some point in their lifetimes; a similar number have such flips
without a major merger a ($58.4\%$, $782$ haloes).  For those that
have a major merger at the same time as such a flip, the fraction is
still very low, at $1.1\%$ ($15$ haloes).


\subsubsection{Flips in haloes without mergers}\label{s:mwlifetimes}
Finally, we consider the particular case of haloes which have quiet merger
histories, i.e. which do not have a major merger after their formation
epoch (i.e. after $z\simeq 1$; see Fig.~\ref{f:MWformtimes}). This
case is interesting because it includes those haloes most likely to
host a disc. 

\begin{figure*} 
  \centering
  \includegraphics[width=\figw]{\fpath 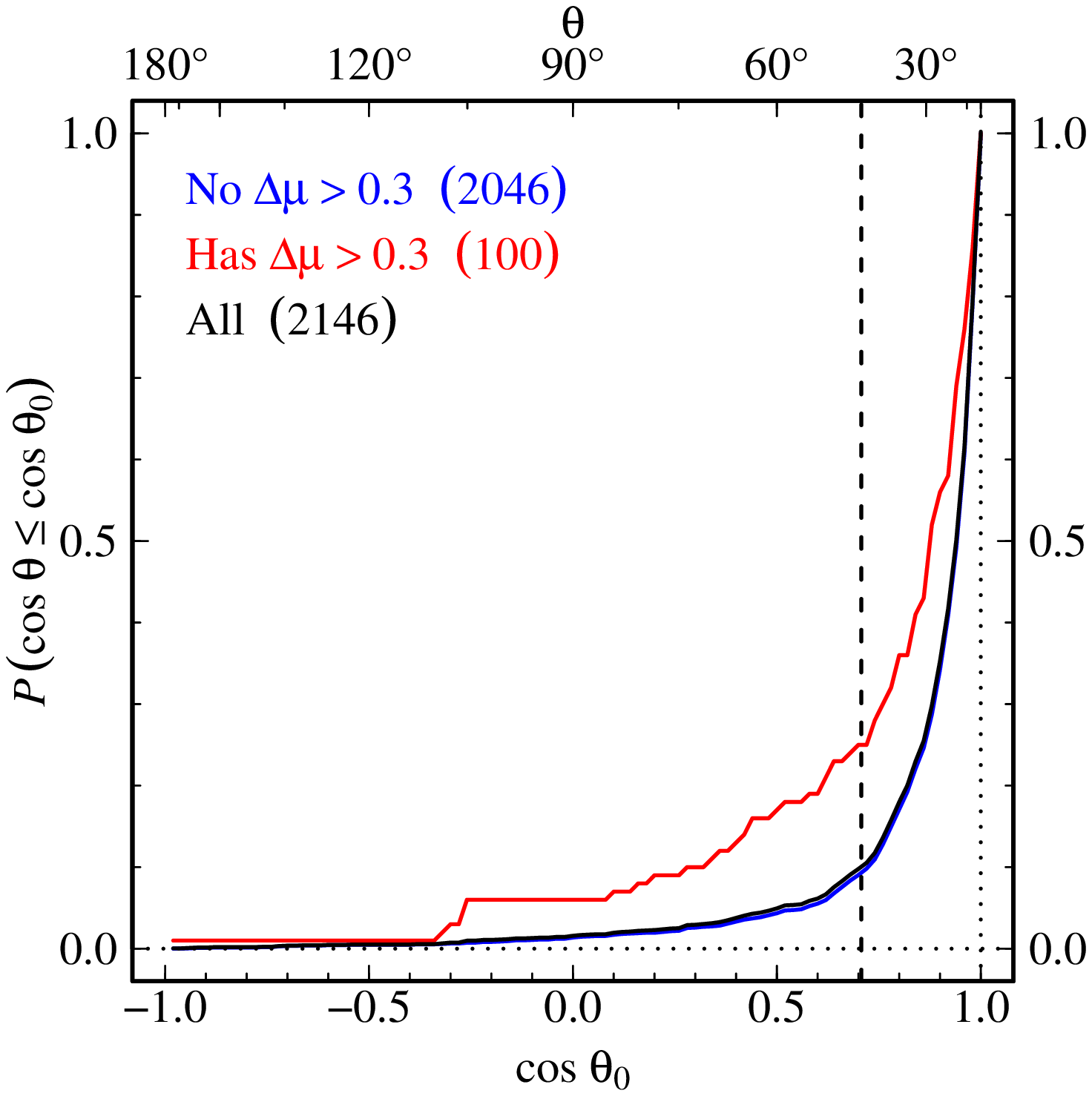}
  \includegraphics[width=\figw]{\fpath 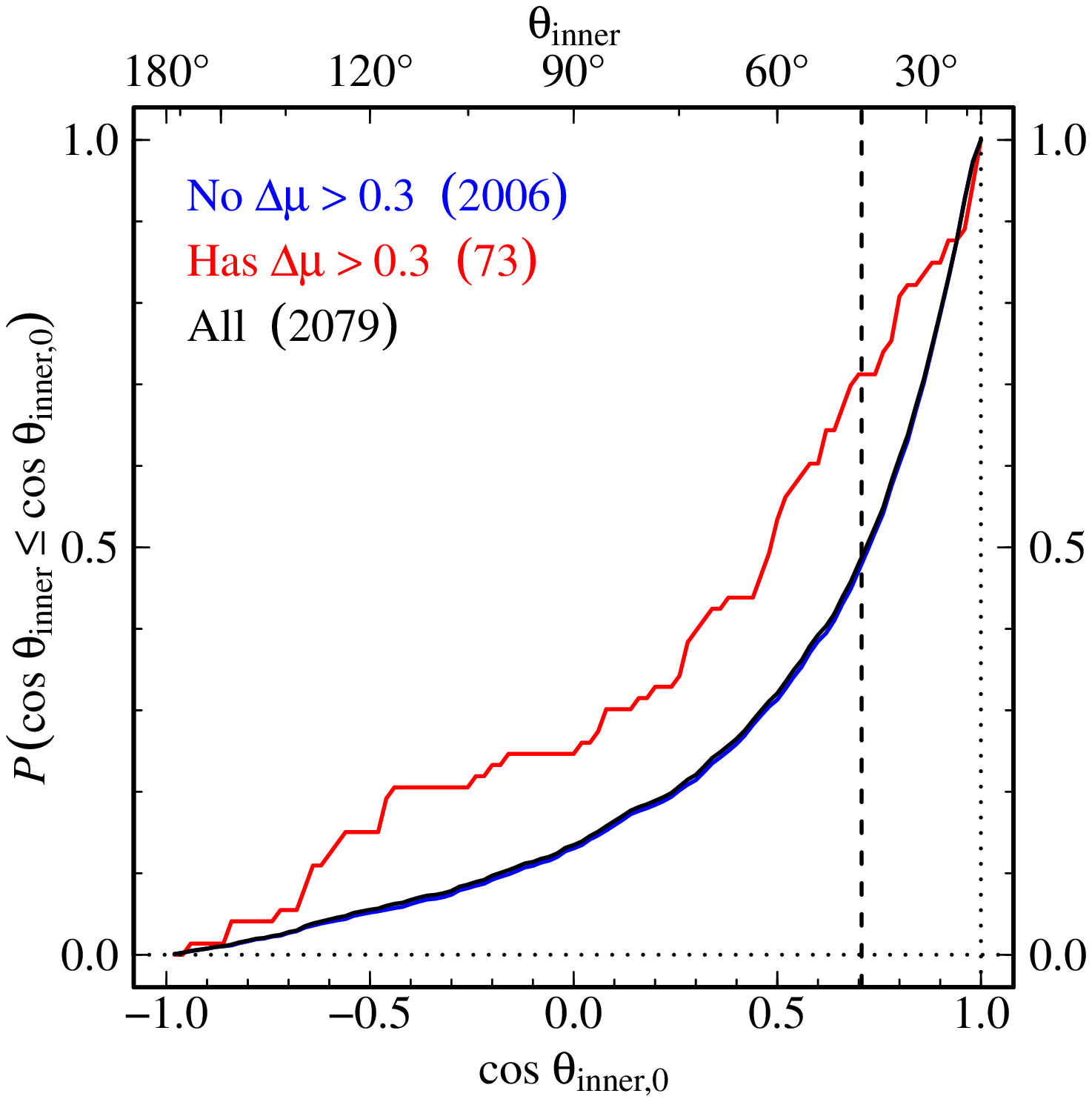}
  \caption{The cumulative fraction of MW final-mass haloes that have a
    spin flip (left: whole halo, right: inner halo) of at least
    $\theta_0$ degrees at some point in their lifetime (black line).
    The other two lines correspond to the cases when the halo does
    (red) or does not (blue) have a major merger ($\DM > 0.3$) during
    its lifetime. (The black line is almost coincident with the blue
    line.)  \pbscripts{fig\_cdf\_cosMWmerghists.R}
    \pbscripts{fig\_cdf\_cosinnerMWmerghists.R}}
  \label{f:mwfracwithflip}
\end{figure*}

The results are shown in Fig.~\ref{f:mwfracwithflip} (left panel).  We
find that $9.0\%$ of haloes without major mergers since formation
nevertheless have a flip of their total spin of at least $45\degr$
($185$ haloes out of $2046$).  Since there are very few major mergers
 after formation \pbnew{even for our
  total halo population} (as shown in Fig.~\ref{f:MWflipdistro}), this
figure does not change much if we do \pbnew{\emph{not}}
apply the no-major-merger restriction: $9.8\%$ of such haloes have
large flips, corresponding to $210$ haloes out of $2146$.  On the
other hand, $25\%$ of the $100$ haloes \emph{with} major mergers since
formation have spin flips of at least $45\degr$.


If we consider flips to the inner halo spin vector (right panel of
Fig.~\ref{f:mwfracwithflip}), then as we have seen before, there is an
increased likelihood for a halo to experience a significant flip.  For
haloes without major mergers after formation, $47\%$ have large flips
of their inner spin ($946$ out of $2006$).  Since large spin flips,
particularly of the inner halo, are so common during the lifetimes of
Milky Way-mass haloes, it seems unlikely that all such flips will
result in the eventual destruction of a disc, although some form of
dynamical disturbance is to be expected. The effect of the flips on
the structure of the disc cannot, of course, be determined with our
dark matter only simulations.



\section{Discussion and conclusions}\label{s:MWconc}
We have investigated the idea that spin flips -- large and rapid
changes in the orientation of the angular momentum of dark matter
haloes -- can occur without \pbnew{a major halo merger. These flips} are a manifestation of strong tidal interactions and can be caused by
minor mergers or by flybys of a neighbouring object. Spin flips could,
in principle, cause enough of a dynamical disturbance in the halo to
disrupt or even destroy a galactic disc, \pbnew{perhaps resulting in
  the formation of a bulge or spheroid}. Evidence for such dramatic
outcomes have been seen in simulations of the formation of individual
galaxies \citep[e.g.][]{2005MNRAS.363.1299O,2009MNRAS.396..696S}.
However, semi-analytic galaxy formation models do not take into
account the potentially destructive effects of spin flips. The only
processes that can transform discs into spheroids in current models
are major mergers and disc instabilities caused by the accretion of
matter onto the disc.

Our goal in this paper has been to determine the frequency of spin
flips during the lifetime of a galactic halo. We have distinguished
between flips affecting the entire halo and flips affecting only the
inner parts of the halo which, at face value, would seem the most
relevant for the stability of the disc.  We have, for this initial
exploration, chosen to focus on haloes that are roughly the mass of
the Milky Way's halo at $z=0$, i.e. between $10^{12}$ and
$10^{12.5}\munit$ \pbnew{and that are reasonably relaxed, as would be
  expected for halos in which discs can form.} In \pbnew{\citetalias{paperII}} we will extend this analysis to
haloes of a much larger range in mass.

We have found that, while the majority of what we have termed
``events'' (i.e. changes to a halo between a given snapshot $t_i$ and
a preceding time $t_i-\tau$) cause only small variations in both mass
and spin direction, the distribution has a significant scatter and a
large tail of \pbnew{significant} variations in spin
direction.  The vast majority of large spin flips affecting the whole
halo occur without an accompanying major merger ($93$ per cent of
events with angular change in spin direction $\theta>45\degr$ have a
fractional mass change $\DM \le 0.3$).  However, such large halo-wide
spin flips are a rare occurrence: only $0.7$ per cent of
non-major-merger events ($\DM \le 0.3$) have $\theta>45\degr$.

Over the course of their lifetime (i.e. \pbnew{over} the period after
the halo has acquired half of its final mass), we find that $10.5$ per
cent of MW final-mass haloes experience at least one spin flip of
$\theta > 45\degr$ with a timescale of $\tau = 0.5\Gyr$; $10.1$ per
cent of the haloes experience such a flip without it coinciding with a
major merger.  These percentages increase for longer timescales and
smaller minimum angle change. Finally, we find that $9$ per cent of
the haloes that have not had \emph{any} major merger after formation
nevertheless have at least one spin flip of $45\degr$ or more.

The spin of the inner halo is subject to \emph{larger} and \emph{more
  frequent} changes in direction than the total halo spin, but like
the total halo spin, inner spin flips also occur mainly without an
accompanying major merger.  Over half of the haloes have large inner
spin flips at some point in their lifetimes without these coinciding
with a major merger.  For the haloes that do not experience any major
mergers after formation, $47\%$ experience a large inner halo spin
flip.  Large spin flips occur sufficiently frequently that they could
have a significant impact on the evolution of the galactic baryonic
material.

Our results suggest that a more complete understanding of the
stability and resilience of galactic discs will require looking beyond
mergers (major or minor) and internal instabilites and should include
the role of spin flips which, as we have seen, can be quite common for
the inner halo. The survivability of discs is not just determined by
the halo potential.  As has been shown in both models
\citep[e.g.][]{2009ApJ...702..307S, 2009ApJ...691.1168H,
  2009MNRAS.397..802H} and simulations
\citep[e.g.][]{2005MNRAS.363.1299O, 2009MNRAS.396..696S}, the details
of the baryonic physics -- gas fraction, strength of supernova
feedback, and other types of interaction between stars and gas -- play
a major role in whether a galactic disc lives or dies, or even can
reform afterwards.  Nevertheless, the behaviour of the underlying dark
matter plays a critical role in galaxy formation. To understand how
spin flips influence the evolution of discs will require full baryon
physics simulations at high resolution. The handful of examples of
simulations we have mentioned here already demonstrate that this
process is both important and tractable.

\section*{Acknowledgements}
\pbnew{PEB thanks Peter Schneider \& Christiano Porciani for helpful
  discussions, and acknowledges the support of the Deutsche
  Forschungsgemeinschaft under the project SCHN 342/7--1 in the
  framework of the Priority Programme SPP-1177, and the Initiative and
  Networking Fund of the Helmholtz Association, contract HA-101
  (``Physics at the Terascale''). }  CSF acknowledges a Royal Societ
Wolfson Research Merit Award \pbnew{and an ERC Advanced Investigator
  grant.}  The simulations and analyses used in this paper were
carried out as part of the programme of the Virgo Consortium on the
Regatta supercomputer of the Computing Centre of the
Max-Planck-Society in Garching, and the Cosmology Machine
supercomputer at the Institute for Computational Cosmology, Durham.
\pbnew{The Cosmology Machine is part of the DiRAC Facility jointly
  funded by STFC, the Large Facilities Capital Fund of BIS, and Durham
  University.  }


\bibliographystyle{mn2eimproved}
\bibliography{spinflips_both}

\begin{thebibliography}{110}
\expandafter\ifx\csname natexlab\endcsname\relax\def\natexlab#1{#1}\fi

\bibitem[{{Agustsson} \& {Brainerd}(2010)}]{2010ApJ...709.1321A}
{Agustsson} I., {Brainerd} T.~G., 2010, \apj, 709, 1321

\bibitem[{{Allgood} {et~al}\mbox{.}(2006){Allgood}, {Flores}, {Primack},
  {Kravtsov}, {Wechsler}, {Faltenbacher}, \& {Bullock}}]{2006MNRAS.367.1781A}
{Allgood} B., {Flores} R.~A., {Primack} J.~R., {Kravtsov} A.~V., {Wechsler}
  R.~H., {Faltenbacher} A., {Bullock} J.~S., 2006, \mnras, 367, 1781

\bibitem[{{Avila-Reese} {et~al}\mbox{.}(2005){Avila-Reese}, {Col{\'{\i}}n},
  {Gottl{\"o}ber}, {Firmani}, \& {Maulbetsch}}]{2005ApJ...634...51A}
{Avila-Reese} V., {Col{\'{\i}}n} P., {Gottl{\"o}ber} S., {Firmani} C.,
  {Maulbetsch} C., 2005, \apj, 634, 51

\bibitem[{{Bailin} \& {Steinmetz}(2005)}]{2005ApJ...627..647B}
{Bailin} J., {Steinmetz} M., 2005, \apj, 627, 647

\bibitem[{{Barnes} \& {Efstathiou}(1987)}]{1987ApJ...319..575B}
{Barnes} J., {Efstathiou} G., 1987, \apj, 319, 575

\bibitem[{{Barnes}(1988)}]{1988ApJ...331..699B}
{Barnes} J.~E., 1988, \apj, 331, 699

\bibitem[{{Barnes}(1992)}]{1992ApJ...393..484B}
---, 1992, \apj, 393, 484

\bibitem[{{Barnes} \& {Hernquist}(1996)}]{1996ApJ...471..115B}
{Barnes} J.~E., {Hernquist} L., 1996, \apj, 471, 115

\bibitem[{{Baugh}(2006)}]{2006RPPh...69.3101B}
{Baugh} C.~M., 2006, Reports on Progress in Physics, 69, 3101

\bibitem[{{Benson}(2010)}]{2010PhR...495...33B}
{Benson} A.~J., 2010, \physrep, 495, 33

\bibitem[{{Bett} {et~al}\mbox{.}(2007){Bett}, {Eke}, {Frenk}, {Jenkins},
  {Helly}, \& {Navarro}}]{bett07}
{Bett} P., {Eke} V., {Frenk} C.~S., {Jenkins} A., {Helly} J., {Navarro} J.,
  2007, \mnras, 376, 215

\bibitem[{{Bett} {et~al}\mbox{.}(2010){Bett}, {Eke}, {Frenk}, {Jenkins}, \&
  {Okamoto}}]{bett10}
{Bett} P., {Eke} V., {Frenk} C.~S., {Jenkins} A., {Okamoto} T., 2010, \mnras,
  404, 1137

\bibitem[{{Bett}(2010)}]{2010AIPC.1240..403B}
{Bett} P.~E., 2010, in American Institute of Physics Conference Series, Vol.
  1240, American Institute of Physics Conference Series, {V.~P.~Debattista \&
  C.~C.~Popescu}, ed., pp. 403--404

\bibitem[{{Bett} \& {Frenk}(2011)}]{paperII}
{Bett} P.~E., {Frenk} C.~S., 2011, (in prep)

\bibitem[{{Bournaud}, {Jog} \& {Combes}(2005){Bournaud}, {Jog}, \&
  {Combes}}]{2005A&A...437...69B}
{Bournaud} F., {Jog} C.~J., {Combes} F., 2005, \aap, 437, 69

\bibitem[{{Bower} {et~al}\mbox{.}(2006){Bower}, {Benson}, {Malbon}, {Helly},
  {Frenk}, {Baugh}, {Cole}, \& {Lacey}}]{bowergalform2006}
{Bower} R.~G., {Benson} A.~J., {Malbon} R., {Helly} J.~C., {Frenk} C.~S.,
  {Baugh} C.~M., {Cole} S., {Lacey} C.~G., 2006, \mnras, 370, 645

\bibitem[{{Boylan-Kolchin} {et~al}\mbox{.}(2009){Boylan-Kolchin}, {Springel},
  {White}, {Jenkins}, \& {Lemson}}]{2009MNRAS.398.1150B}
{Boylan-Kolchin} M., {Springel} V., {White} S.~D.~M., {Jenkins} A., {Lemson}
  G., 2009, \mnras, 398, 1150

\bibitem[{{Brunino} {et~al}\mbox{.}(2007){Brunino}, {Trujillo}, {Pearce}, \&
  {Thomas}}]{2007MNRAS.375..184B}
{Brunino} R., {Trujillo} I., {Pearce} F.~R., {Thomas} P.~A., 2007, \mnras, 375,
  184

\bibitem[{{Bryan} \& {Norman}(1998)}]{BN98}
{Bryan} G.~L., {Norman} M.~L., 1998, \apj, 495, 80

\bibitem[{{Bullock} {et~al}\mbox{.}(2001){Bullock}, {Dekel}, {Kolatt},
  {Kravtsov}, {Klypin}, {Porciani}, \& {Primack}}]{2001ApJ...555..240B}
{Bullock} J.~S., {Dekel} A., {Kolatt} T.~S., {Kravtsov} A.~V., {Klypin} A.~A.,
  {Porciani} C., {Primack} J.~R., 2001, \apj, 555, 240

\bibitem[{{Catelan} \& {Theuns}(1996{\natexlab{a}})}]{1996MNRAS.282..436C}
{Catelan} P., {Theuns} T., 1996{\natexlab{a}}, \mnras, 282, 436

\bibitem[{{Catelan} \& {Theuns}(1996{\natexlab{b}})}]{1996MNRAS.282..455C}
---, 1996{\natexlab{b}}, \mnras, 282, 455

\bibitem[{{Chen}, {Jing} \& {Yoshikawa}(2003){Chen}, {Jing}, \&
  {Yoshikawa}}]{2003ApJ...597...35C}
{Chen} D.~N., {Jing} Y.~P., {Yoshikawa} K., 2003, \apj, 597, 35

\bibitem[{{Cole} \& {Lacey}(1996)}]{1996MNRAS.281..716C}
{Cole} S., {Lacey} C., 1996, \mnras, 281, 716

\bibitem[{{Colless} {et~al}\mbox{.}(2001){Colless}, {Dalton}, {Maddox},
  {Sutherland}, {Norberg}, {Cole}, {Bland-Hawthorn}, {Bridges}, {Cannon},
  {Collins}, {Couch}, {Cross}, {Deeley}, {De Propris}, {Driver}, {Efstathiou},
  {Ellis}, {Frenk}, {Glazebrook}, {Jackson}, {Lahav}, {Lewis}, {Lumsden},
  {Madgwick}, {Peacock}, {Peterson}, {Price}, {Seaborne}, \&
  {Taylor}}]{2001MNRAS.328.1039C}
{Colless} M. {et~al.}, 2001, \mnras, 328, 1039

\bibitem[{{Cox} {et~al}\mbox{.}(2006){Cox}, {Dutta}, {Di Matteo}, {Hernquist},
  {Hopkins}, {Robertson}, \& {Springel}}]{2006ApJ...650..791C}
{Cox} T.~J., {Dutta} S.~N., {Di Matteo} T., {Hernquist} L., {Hopkins} P.~F.,
  {Robertson} B., {Springel} V., 2006, \apj, 650, 791

\bibitem[{{Cox} {et~al}\mbox{.}(2008){Cox}, {Jonsson}, {Somerville}, {Primack},
  \& {Dekel}}]{2008MNRAS.384..386C}
{Cox} T.~J., {Jonsson} P., {Somerville} R.~S., {Primack} J.~R., {Dekel} A.,
  2008, \mnras, 384, 386

\bibitem[{{Croft} {et~al}\mbox{.}(2009){Croft}, {Di Matteo}, {Springel}, \&
  {Hernquist}}]{2009MNRAS.400...43C}
{Croft} R.~A.~C., {Di Matteo} T., {Springel} V., {Hernquist} L., 2009, \mnras,
  400, 43

\bibitem[{{Cuesta} {et~al}\mbox{.}(2008){Cuesta}, {Betancort-Rijo},
  {Gottl{\"o}ber}, {Patiri}, {Yepes}, \& {Prada}}]{2008MNRAS.385..867C}
{Cuesta} A.~J., {Betancort-Rijo} J.~E., {Gottl{\"o}ber} S., {Patiri} S.~G.,
  {Yepes} G., {Prada} F., 2008, \mnras, 385, 867

\bibitem[{{Davis} {et~al}\mbox{.}(1985){Davis}, {Efstathiou}, {Frenk}, \&
  {White}}]{defw85}
{Davis} M., {Efstathiou} G., {Frenk} C.~S., {White} S.~D.~M., 1985, \apj, 292,
  371

\bibitem[{{De Lucia} \& {Blaizot}(2007)}]{2007MNRAS.375....2D}
{De Lucia} G., {Blaizot} J., 2007, \mnras, 375, 2

\bibitem[{{De Lucia} {et~al}\mbox{.}(2010){De Lucia}, {Boylan-Kolchin},
  {Benson}, {Fontanot}, \& {Monaco}}]{2010MNRAS.406.1533D}
{De Lucia} G., {Boylan-Kolchin} M., {Benson} A.~J., {Fontanot} F., {Monaco} P.,
  2010, \mnras, 406, 1533

\bibitem[{{Deason} {et~al}\mbox{.}(2011){Deason}, {McCarthy}, {Font}, {Evans},
  {Frenk}, {Belokurov}, {Libeskind}, {Crain}, \&
  {Theuns}}]{2011MNRAS.415.2607D}
{Deason} A.~J. {et~al.}, 2011, \mnras, 415, 2607

\bibitem[{{Doroshkevich}(1970{\natexlab{a}})}]{1970Ap......6..320D}
{Doroshkevich} A.~G., 1970{\natexlab{a}}, Astrophysics, 6, 320

\bibitem[{{Doroshkevich}(1970{\natexlab{b}})}]{1970Afz.....6..581D}
---, 1970{\natexlab{b}}, Astrofizika, 6, 581

\bibitem[{{Efstathiou} \& {Jones}(1979)}]{1979MNRAS.186..133E}
{Efstathiou} G., {Jones} B.~J.~T., 1979, \mnras, 186, 133

\bibitem[{{Eke}, {Cole} \& {Frenk}(1996){Eke}, {Cole}, \& {Frenk}}]{ECF96}
{Eke} V.~R., {Cole} S., {Frenk} C.~S., 1996, \mnras, 282, 263

\bibitem[{{Fakhouri} \& {Ma}(2008)}]{2008MNRAS.386..577F}
{Fakhouri} O., {Ma} C., 2008, \mnras, 386, 577

\bibitem[{{Fakhouri} \& {Ma}(2009)}]{2009MNRAS.394.1825F}
---, 2009, \mnras, 394, 1825

\bibitem[{{Fall} \& {Efstathiou}(1980)}]{1980MNRAS.193..189F}
{Fall} S.~M., {Efstathiou} G., 1980, \mnras, 193, 189

\bibitem[{{Faltenbacher} {et~al}\mbox{.}(2005){Faltenbacher}, {Allgood},
  {Gottl{\"o}ber}, {Yepes}, \& {Hoffman}}]{2005MNRAS.362.1099F}
{Faltenbacher} A., {Allgood} B., {Gottl{\"o}ber} S., {Yepes} G., {Hoffman} Y.,
  2005, \mnras, 362, 1099

\bibitem[{{Freeman}(2008)}]{2008IAUS..245....3F}
{Freeman} K.~C., 2008, in IAU Symposium, Vol. 245, IAU Symposium, {M.~Bureau,
  E.~Athanassoula, \& B.~Barbuy}, ed., pp. 3--10

\bibitem[{{Frenk} {et~al}\mbox{.}(1988){Frenk}, {White}, {Davis}, \&
  {Efstathiou}}]{1988ApJ...327..507F}
{Frenk} C.~S., {White} S.~D.~M., {Davis} M., {Efstathiou} G., 1988, \apj, 327,
  507

\bibitem[{{Gao} {et~al}\mbox{.}(2008){Gao}, {Navarro}, {Cole}, {Frenk},
  {White}, {Springel}, {Jenkins}, \& {Neto}}]{2008MNRAS.387..536G}
{Gao} L., {Navarro} J.~F., {Cole} S., {Frenk} C.~S., {White} S.~D.~M.,
  {Springel} V., {Jenkins} A., {Neto} A.~F., 2008, \mnras, 387, 536

\bibitem[{{Genel} {et~al}\mbox{.}(2009){Genel}, {Genzel}, {Bouch{\'e}}, {Naab},
  \& {Sternberg}}]{2009ApJ...701.2002G}
{Genel} S., {Genzel} R., {Bouch{\'e}} N., {Naab} T., {Sternberg} A., 2009,
  \apj, 701, 2002

\bibitem[{{Giocoli} {et~al}\mbox{.}(2007){Giocoli}, {Moreno}, {Sheth}, \&
  {Tormen}}]{2007MNRAS.376..977G}
{Giocoli} C., {Moreno} J., {Sheth} R.~K., {Tormen} G., 2007, \mnras, 376, 977

\bibitem[{{Gustafsson}, {Fairbairn} \& {Sommer-Larsen}(2006){Gustafsson},
  {Fairbairn}, \& {Sommer-Larsen}}]{2006PhRvD..74l3522G}
{Gustafsson} M., {Fairbairn} M., {Sommer-Larsen} J., 2006, \prd, 74, 123522

\bibitem[{{Hahn} {et~al}\mbox{.}(2007{\natexlab{a}}){Hahn}, {Carollo},
  {Porciani}, \& {Dekel}}]{2007MNRAS.381...41H}
{Hahn} O., {Carollo} C.~M., {Porciani} C., {Dekel} A., 2007{\natexlab{a}},
  \mnras, 381, 41

\bibitem[{{Hahn} {et~al}\mbox{.}(2007{\natexlab{b}}){Hahn}, {Porciani},
  {Carollo}, \& {Dekel}}]{2007MNRAS.375..489H}
{Hahn} O., {Porciani} C., {Carollo} C.~M., {Dekel} A., 2007{\natexlab{b}},
  \mnras, 375, 489

\bibitem[{{Hahn}, {Teyssier} \& {Carollo}(2010){Hahn}, {Teyssier}, \&
  {Carollo}}]{2010MNRAS.405..274H}
{Hahn} O., {Teyssier} R., {Carollo} C.~M., 2010, \mnras, 405, 274

\bibitem[{{Harker} {et~al}\mbox{.}(2006){Harker}, {Cole}, {Helly}, {Frenk}, \&
  {Jenkins}}]{harker2006}
{Harker} G., {Cole} S., {Helly} J., {Frenk} C., {Jenkins} A., 2006, \mnras,
  367, 1039

\bibitem[{{Hayashi}, {Navarro} \& {Springel}(2007){Hayashi}, {Navarro}, \&
  {Springel}}]{2007MNRAS.377...50H}
{Hayashi} E., {Navarro} J.~F., {Springel} V., 2007, \mnras, 377, 50

\bibitem[{{Helly} {et~al}\mbox{.}(2003){Helly}, {Cole}, {Frenk}, {Baugh},
  {Benson}, \& {Lacey}}]{helly2003}
{Helly} J.~C., {Cole} S., {Frenk} C.~S., {Baugh} C.~M., {Benson} A., {Lacey}
  C., 2003, \mnras, 338, 903

\bibitem[{{Hernquist}(1992)}]{1992ApJ...400..460H}
{Hernquist} L., 1992, \apj, 400, 460

\bibitem[{{Hernquist}(1993)}]{1993ApJ...409..548H}
---, 1993, \apj, 409, 548

\bibitem[{{Hopkins} {et~al}\mbox{.}(2010){Hopkins}, {Bundy}, {Croton},
  {Hernquist}, {Keres}, {Khochfar}, {Stewart}, {Wetzel}, \&
  {Younger}}]{2010ApJ...715..202H}
{Hopkins} P.~F. {et~al.}, 2010, \apj, 715, 202

\bibitem[{{Hopkins} {et~al}\mbox{.}(2009{\natexlab{a}}){Hopkins}, {Cox},
  {Younger}, \& {Hernquist}}]{2009ApJ...691.1168H}
{Hopkins} P.~F., {Cox} T.~J., {Younger} J.~D., {Hernquist} L.,
  2009{\natexlab{a}}, \apj, 691, 1168

\bibitem[{{Hopkins} {et~al}\mbox{.}(2009{\natexlab{b}}){Hopkins}, {Somerville},
  {Cox}, {Hernquist}, {Jogee}, {Kere{\v s}}, {Ma}, {Robertson}, \&
  {Stewart}}]{2009MNRAS.397..802H}
{Hopkins} P.~F. {et~al.}, 2009{\natexlab{b}}, \mnras, 397, 802

\bibitem[{{Hoyle}(1951)}]{1951pca..conf..195H}
{Hoyle} F., 1951, in Problems of Cosmical Aerodynamics, {Burgers} J.~M., {van
  de Hulst} H.~C., eds., Central Air Documents Office, Dayton, OH, pp. 195--197

\bibitem[{{Knebe} {et~al}\mbox{.}(2004){Knebe}, {Gill}, {Gibson}, {Lewis},
  {Ibata}, \& {Dopita}}]{2004ApJ...603....7K}
{Knebe} A., {Gill} S.~P.~D., {Gibson} B.~K., {Lewis} G.~F., {Ibata} R.~A.,
  {Dopita} M.~A., 2004, \apj, 603, 7

\bibitem[{{Knebe} \& {Power}(2008)}]{2008ApJ...678..621K}
{Knebe} A., {Power} C., 2008, \apj, 678, 621

\bibitem[{{Kormendy} \& {Kennicutt}(2004)}]{2004ARA&A..42..603K}
{Kormendy} J., {Kennicutt}, Jr. R.~C., 2004, \araa, 42, 603

\bibitem[{{Lacey} \& {Cole}(1993)}]{1993MNRAS.262..627L}
{Lacey} C., {Cole} S., 1993, \mnras, 262, 627

\bibitem[{{Lemson} \& {the Virgo Consortium}(2006)}]{milldb}
{Lemson} G., {the Virgo Consortium}, 2006, ArXiv Astrophysics e-prints
  (astro-ph/0608019)

\bibitem[{{Li}, {Mo} \& {Gao}(2008){Li}, {Mo}, \& {Gao}}]{2008MNRAS.389.1419L}
{Li} Y., {Mo} H.~J., {Gao} L., 2008, \mnras, 389, 1419

\bibitem[{{Libeskind} {et~al}\mbox{.}(2005){Libeskind}, {Frenk}, {Cole},
  {Helly}, {Jenkins}, {Navarro}, \& {Power}}]{2005MNRAS.363..146L}
{Libeskind} N.~I., {Frenk} C.~S., {Cole} S., {Helly} J.~C., {Jenkins} A.,
  {Navarro} J.~F., {Power} C., 2005, \mnras, 363, 146

\bibitem[{{Libeskind} {et~al}\mbox{.}(2009){Libeskind}, {Frenk}, {Cole},
  {Jenkins}, \& {Helly}}]{2009MNRAS.399..550L}
{Libeskind} N.~I., {Frenk} C.~S., {Cole} S., {Jenkins} A., {Helly} J.~C., 2009,
  \mnras, 399, 550

\bibitem[{{Libeskind} {et~al}\mbox{.}(2011){Libeskind}, {Knebe}, {Hoffman},
  {Gottl{\"o}ber}, {Yepes}, \& {Steinmetz}}]{2011MNRAS.411.1525L}
{Libeskind} N.~I., {Knebe} A., {Hoffman} Y., {Gottl{\"o}ber} S., {Yepes} G.,
  {Steinmetz} M., 2011, \mnras, 411, 1525

\bibitem[{{{\L}okas} \& {Mamon}(2001)}]{2001MNRAS.321..155L}
{{\L}okas} E.~L., {Mamon} G.~A., 2001, \mnras, 321, 155

\bibitem[{{Lovell} {et~al}\mbox{.}(2011){Lovell}, {Eke}, {Frenk}, \&
  {Jenkins}}]{2011MNRAS.413.3013L}
{Lovell} M.~R., {Eke} V.~R., {Frenk} C.~S., {Jenkins} A., 2011, \mnras, 413,
  3013

\bibitem[{{Macci{\`o}}, {Dutton} \& {van den Bosch}(2008){Macci{\`o}},
  {Dutton}, \& {van den Bosch}}]{Maccio2008}
{Macci{\`o}} A.~V., {Dutton} A.~A., {van den Bosch} F.~C., 2008, \mnras, 391,
  1940

\bibitem[{{Macci{\`o}} {et~al}\mbox{.}(2007){Macci{\`o}}, {Dutton}, {van den
  Bosch}, {Moore}, {Potter}, \& {Stadel}}]{Maccio2007}
{Macci{\`o}} A.~V., {Dutton} A.~A., {van den Bosch} F.~C., {Moore} B., {Potter}
  D., {Stadel} J., 2007, \mnras, 378, 55

\bibitem[{{Maller} {et~al}\mbox{.}(2006){Maller}, {Katz}, {Kere{\v s}},
  {Dav{\'e}}, \& {Weinberg}}]{2006ApJ...647..763M}
{Maller} A.~H., {Katz} N., {Kere{\v s}} D., {Dav{\'e}} R., {Weinberg} D.~H.,
  2006, \apj, 647, 763

\bibitem[{{Mo}, {Mao} \& {White}(1998){Mo}, {Mao}, \&
  {White}}]{1998MNRAS.295..319M}
{Mo} H.~J., {Mao} S., {White} S.~D.~M., 1998, \mnras, 295, 319

\bibitem[{{Mu{\~n}oz-Cuartas} {et~al}\mbox{.}(2011){Mu{\~n}oz-Cuartas},
  {Macci{\`o}}, {Gottl{\"o}ber}, \& {Dutton}}]{2011MNRAS.411..584M}
{Mu{\~n}oz-Cuartas} J.~C., {Macci{\`o}} A.~V., {Gottl{\"o}ber} S., {Dutton}
  A.~A., 2011, \mnras, 411, 584

\bibitem[{{Naab} \& {Burkert}(2003)}]{2003ApJ...597..893N}
{Naab} T., {Burkert} A., 2003, \apj, 597, 893

\bibitem[{{Navarro}, {Frenk} \& {White}(1996){Navarro}, {Frenk}, \&
  {White}}]{nfw96}
{Navarro} J.~F., {Frenk} C.~S., {White} S.~D.~M., 1996, \apj, 462, 563

\bibitem[{{Navarro}, {Frenk} \& {White}(1997){Navarro}, {Frenk}, \&
  {White}}]{nfw97}
---, 1997, \apj, 490, 493

\bibitem[{{Neto} {et~al}\mbox{.}(2007){Neto}, {Gao}, {Bett}, {Cole}, {Navarro},
  {Frenk}, {White}, {Springel}, \& {Jenkins}}]{neto07}
{Neto} A.~F. {et~al.}, 2007, \mnras, 381, 1450

\bibitem[{{Okamoto} {et~al}\mbox{.}(2005){Okamoto}, {Eke}, {Frenk}, \&
  {Jenkins}}]{2005MNRAS.363.1299O}
{Okamoto} T., {Eke} V.~R., {Frenk} C.~S., {Jenkins} A., 2005, \mnras, 363, 1299

\bibitem[{{Ostriker} \& {Binney}(1989)}]{1989MNRAS.237..785O}
{Ostriker} E.~C., {Binney} J.~J., 1989, \mnras, 237, 785

\bibitem[{{Parry}, {Eke} \& {Frenk}(2009){Parry}, {Eke}, \& {Frenk}}]{Parry09}
{Parry} O.~H., {Eke} V.~R., {Frenk} C.~S., 2009, \mnras, 396, 1972

\bibitem[{{Paz}, {Stasyszyn} \& {Padilla}(2008){Paz}, {Stasyszyn}, \&
  {Padilla}}]{2008MNRAS.389.1127P}
{Paz} D.~J., {Stasyszyn} F., {Padilla} N.~D., 2008, \mnras, 389, 1127

\bibitem[{{Peebles}(1969)}]{1969ApJ...155..393P}
{Peebles} P.~J.~E., 1969, \apj, 155, 393

\bibitem[{{Peebles}(1971)}]{1971A&A....11..377P}
---, 1971, \aap, 11, 377

\bibitem[{{Percival} {et~al}\mbox{.}(2002){Percival}, {Sutherland}, {Peacock},
  {Baugh}, {Bland-Hawthorn}, {Bridges}, {Cannon}, {Cole}, {Colless}, {Collins},
  {Couch}, {Dalton}, {De Propris}, {Driver}, {Efstathiou}, {Ellis}, {Frenk},
  {Glazebrook}, {Jackson}, {Lahav}, {Lewis}, {Lumsden}, {Maddox}, {Moody},
  {Norberg}, {Peterson}, \& {Taylor}}]{2002MNRAS.337.1068P}
{Percival} W.~J. {et~al.}, 2002, \mnras, 337, 1068

\bibitem[{{Porciani}, {Dekel} \& {Hoffman}(2002){Porciani}, {Dekel}, \&
  {Hoffman}}]{2002MNRAS.332..325P}
{Porciani} C., {Dekel} A., {Hoffman} Y., 2002, \mnras, 332, 325

\bibitem[{{Romano-D{\'{\i}}az} {et~al}\mbox{.}(2009){Romano-D{\'{\i}}az},
  {Shlosman}, {Heller}, \& {Hoffman}}]{2009ApJ...702.1250R}
{Romano-D{\'{\i}}az} E., {Shlosman} I., {Heller} C., {Hoffman} Y., 2009, \apj,
  702, 1250

\bibitem[{{Scannapieco} {et~al}\mbox{.}(2009){Scannapieco}, {White},
  {Springel}, \& {Tissera}}]{2009MNRAS.396..696S}
{Scannapieco} C., {White} S.~D.~M., {Springel} V., {Tissera} P.~B., 2009,
  \mnras, 396, 696

\bibitem[{{Sch{\"a}fer}(2009)}]{2009IJMPD..18..173S}
{Sch{\"a}fer} B.~M., 2009, International Journal of Modern Physics D, 18, 173

\bibitem[{{Shaw} {et~al}\mbox{.}(2006){Shaw}, {Weller}, {Ostriker}, \&
  {Bode}}]{2006ApJ...646..815S}
{Shaw} L.~D., {Weller} J., {Ostriker} J.~P., {Bode} P., 2006, \apj, 646, 815

\bibitem[{{Sheth} \& {Tormen}(2004)}]{2004MNRAS.349.1464S}
{Sheth} R.~K., {Tormen} G., 2004, \mnras, 349, 1464

\bibitem[{{Sinha} \& {Holley-Bockelmann}(2011)}]{2011arXiv1103.1675S}
{Sinha} M., {Holley-Bockelmann} K., 2011, \apj, submitted (arXiv:1103.1675)

\bibitem[{{Spergel} {et~al}\mbox{.}(2003){Spergel}, {Verde}, {Peiris},
  {Komatsu}, {Nolta}, {Bennett}, {Halpern}, {Hinshaw}, {Jarosik}, {Kogut},
  {Limon}, {Meyer}, {Page}, {Tucker}, {Weiland}, {Wollack}, \&
  {Wright}}]{wmap1cos2003}
{Spergel} D.~N. {et~al.}, 2003, \apjs, 148, 175

\bibitem[{{Springel} {et~al}\mbox{.}(2005){Springel}, {White}, {Jenkins},
  {Frenk}, {Yoshida}, {Gao}, {Navarro}, {Thacker}, {Croton}, {Helly},
  {Peacock}, {Cole}, {Thomas}, {Couchman}, {Evrard}, {Colberg}, \&
  {Pearce}}]{mill2005}
{Springel} V. {et~al.}, 2005, \nat, 435, 629

\bibitem[{{Springel} {et~al}\mbox{.}(2001){Springel}, {White}, {Tormen}, \&
  {Kauffmann}}]{subfind2001}
{Springel} V., {White} S.~D.~M., {Tormen} G., {Kauffmann} G., 2001, \mnras,
  328, 726

\bibitem[{{Stewart} {et~al}\mbox{.}(2009){Stewart}, {Bullock}, {Wechsler}, \&
  {Maller}}]{2009ApJ...702..307S}
{Stewart} K.~R., {Bullock} J.~S., {Wechsler} R.~H., {Maller} A.~H., 2009, \apj,
  702, 307

\bibitem[{{Stewart} {et~al}\mbox{.}(2008){Stewart}, {Bullock}, {Wechsler},
  {Maller}, \& {Zentner}}]{2008ApJ...683..597S}
{Stewart} K.~R., {Bullock} J.~S., {Wechsler} R.~H., {Maller} A.~H., {Zentner}
  A.~R., 2008, \apj, 683, 597

\bibitem[{{Sugerman}, {Summers} \& {Kamionkowski}(2000){Sugerman}, {Summers},
  \& {Kamionkowski}}]{2000MNRAS.311..762S}
{Sugerman} B., {Summers} F.~J., {Kamionkowski} M., 2000, \mnras, 311, 762

\bibitem[{{Toomre}(1977)}]{1977egsp.conf..401T}
{Toomre} A., 1977, in Evolution of Galaxies and Stellar Populations,
  {B.~M.~Tinsley \& R.~B.~Larson}, ed., pp. 401--416

\bibitem[{{Toomre} \& {Toomre}(1972)}]{1972ApJ...178..623T}
{Toomre} A., {Toomre} J., 1972, \apj, 178, 623

\bibitem[{{Tweed} {et~al}\mbox{.}(2009){Tweed}, {Devriendt}, {Blaizot},
  {Colombi}, \& {Slyz}}]{2009A&A...506..647T}
{Tweed} D., {Devriendt} J., {Blaizot} J., {Colombi} S., {Slyz} A., 2009, \aap,
  506, 647

\bibitem[{{van den Bosch} {et~al}\mbox{.}(2002){van den Bosch}, {Abel},
  {Croft}, {Hernquist}, \& {White}}]{2002ApJ...576...21V}
{van den Bosch} F.~C., {Abel} T., {Croft} R.~A.~C., {Hernquist} L., {White}
  S.~D.~M., 2002, \apj, 576, 21

\bibitem[{{van den Bosch}, {Abel} \& {Hernquist}(2003){van den Bosch}, {Abel},
  \& {Hernquist}}]{2003MNRAS.346..177V}
{van den Bosch} F.~C., {Abel} T., {Hernquist} L., 2003, \mnras, 346, 177

\bibitem[{{Warren} {et~al}\mbox{.}(1992){Warren}, {Quinn}, {Salmon}, \&
  {Zurek}}]{1992ApJ...399..405W}
{Warren} M.~S., {Quinn} P.~J., {Salmon} J.~K., {Zurek} W.~H., 1992, \apj, 399,
  405

\bibitem[{{Wechsler} {et~al}\mbox{.}(2002){Wechsler}, {Bullock}, {Primack},
  {Kravtsov}, \& {Dekel}}]{2002ApJ...568...52W}
{Wechsler} R.~H., {Bullock} J.~S., {Primack} J.~R., {Kravtsov} A.~V., {Dekel}
  A., 2002, \apj, 568, 52

\bibitem[{{White}(1984)}]{1984ApJ...286...38W}
{White} S.~D.~M., 1984, \apj, 286, 38

\bibitem[{{White} \& {Frenk}(1991)}]{1991ApJ...379...52W}
{White} S.~D.~M., {Frenk} C.~S., 1991, \apj, 379, 52

\bibitem[{{White} \& {Rees}(1978)}]{whiterees1978}
{White} S.~D.~M., {Rees} M.~J., 1978, \mnras, 183, 341

\bibitem[{{Zavala}, {Okamoto} \& {Frenk}(2008){Zavala}, {Okamoto}, \&
  {Frenk}}]{Zavala_Frenk08}
{Zavala} J., {Okamoto} T., {Frenk} C.~S., 2008, \mnras, 387, 364

\end{thebibliography}



\label{lastpage}
\end{document}